\documentclass[aps,prx,amsmath,amssymb,superscriptaddress,twocolumn,notitlepage,nofootinbib,longbibliography]{revtex4-2}
\pdfoutput=1

\usepackage[final]{graphicx}
\usepackage{times,bbm,amsmath,amssymb,amsthm}
\usepackage{epsfig,color}
\usepackage[dvipsnames]{xcolor}
\usepackage[colorlinks=true,citecolor=blue,linkcolor=blue]{hyperref}
\hypersetup{
    allcolors  = {blue},
}

\usepackage{cleveref}

\usepackage{float,siunitx}
\usepackage[caption = false]{subfig}
\usepackage{verbatim}
\usepackage[greek,english]{babel}
\usepackage{thumbpdf,enumerate}
\usepackage{booktabs}
\usepackage{sidecap}
\usepackage[scaled=.8]{couriers}    
\usepackage{pstricks}
\usepackage{multirow}
\usepackage{placeins}
\usepackage{relsize}  
\usepackage{pst-grad,bm}
\usepackage{epigraph}
\usepackage{gensymb}
\usepackage{longtable}
\usepackage{booktabs}
\usepackage{gensymb}

\usepackage{soul}
\usepackage{ulem} 
\usepackage{acronym}
\usepackage{physics}
\usepackage{tikz}

\theoremstyle{plain}

\newtheorem*{theorem*}    {Theorem}
\newtheorem*{proposition*}{Proposition}
\newtheorem*{lemma*}      {Lemma}
\newtheorem*{corollary*}  {Corollary}
\newtheorem*{conjecture*} {Conjecture}

\graphicspath{{Images/}}

\begin{document}
\title{Semi-device independent characterization of multiphoton indistinguishability}

\author{Giovanni Rodari}
\thanks{These two authors contributed equally}
\affiliation{Dipartimento di Fisica, Sapienza Universit\`{a} di Roma,
Piazzale Aldo Moro 5, I-00185 Roma, Italy
}

\author{Leonardo Novo}
\thanks{These two authors contributed equally}
\affiliation{International Iberian Nanotechnology Laboratory (INL)
 Av. Mestre José Veiga s/n, 4715-330 Braga, Portugal
}

\author{Riccardo Albiero}
\affiliation{Dipartimento di Fisica, Politecnico di Milano,
Piazza Leonardo da Vinci 32, I-20133 Milano, Italy}
\affiliation{Istituto di Fotonica e Nanotecnologie, Consiglio Nazionale delle Ricerche (IFN-CNR), 
Piazza Leonardo da Vinci 32, I-20133 Milano, Italy
}

\author{Alessia Suprano}
\affiliation{Dipartimento di Fisica, Sapienza Universit\`{a} di Roma,
Piazzale Aldo Moro 5, I-00185 Roma, Italy
}

\author{Carlos T. Tavares}
\affiliation{International Iberian Nanotechnology Laboratory (INL)
 Av. Mestre José Veiga s/n, 4715-330 Braga, Portugal
}

\author{Eugenio Caruccio}
\affiliation{Dipartimento di Fisica, Sapienza Universit\`{a} di Roma,
Piazzale Aldo Moro 5, I-00185 Roma, Italy
}

\author{Francesco Hoch}
\affiliation{Dipartimento di Fisica, Sapienza Universit\`{a} di Roma,
Piazzale Aldo Moro 5, I-00185 Roma, Italy
}

\author{Taira Giordani}
\affiliation{Dipartimento di Fisica, Sapienza Universit\`{a} di Roma,
Piazzale Aldo Moro 5, I-00185 Roma, Italy
}

\author{Gonzalo Carvacho}
\affiliation{Dipartimento di Fisica, Sapienza Universit\`{a} di Roma,
Piazzale Aldo Moro 5, I-00185 Roma, Italy
}

\author{Marco Gardina}
\affiliation{Ephos, Inc., 
333 Bush Street, San Francisco, CA 94104, USA
}

\author{Niki Di Giano}
\affiliation{Dipartimento di Fisica, Politecnico di Milano,
Piazza Leonardo da Vinci 32, I-20133 Milano, Italy}
\affiliation{Istituto di Fotonica e Nanotecnologie, Consiglio Nazionale delle Ricerche (IFN-CNR), 
Piazza Leonardo da Vinci 32, I-20133 Milano, Italy
}

\author{Serena Di Giorgio}
\affiliation{Ephos, Inc., 
333 Bush Street, San Francisco, CA 94104, USA
}

\author{Giacomo Corrielli}
\affiliation{Istituto di Fotonica e Nanotecnologie, Consiglio Nazionale delle Ricerche (IFN-CNR), 
Piazza Leonardo da Vinci 32, I-20133 Milano, Italy
}
\affiliation{Ephos, Inc., 
333 Bush Street, San Francisco, CA 94104, USA
}

\author{Francesco Ceccarelli}
\affiliation{Istituto di Fotonica e Nanotecnologie, Consiglio Nazionale delle Ricerche (IFN-CNR), 
Piazza Leonardo da Vinci 32, I-20133 Milano, Italy
}
\affiliation{Ephos, Inc., 
333 Bush Street, San Francisco, CA 94104, USA
}

\author{Roberto Osellame}
\email{roberto.osellame@cnr.it}
\affiliation{Istituto di Fotonica e Nanotecnologie, Consiglio Nazionale delle Ricerche (IFN-CNR), 
Piazza Leonardo da Vinci 32, I-20133 Milano, Italy
}
\affiliation{Ephos, Inc., 
333 Bush Street, San Francisco, CA 94104, USA
}

\author{Nicol\`o Spagnolo}
\email{nicolo.spagnolo@uniroma1.it}
\affiliation{Dipartimento di Fisica, Sapienza Universit\`{a} di Roma,
Piazzale Aldo Moro 5, I-00185 Roma, Italy
}

\author{Ernesto F. Galv\~{a}o}
\email{ernesto.galvao@inl.int}
\affiliation{International Iberian Nanotechnology Laboratory (INL)
 Av. Mestre José Veiga s/n, 4715-330 Braga, Portugal
}

\author{Fabio Sciarrino}
\affiliation{Dipartimento di Fisica, Sapienza Universit\`{a} di Roma,
Piazzale Aldo Moro 5, I-00185 Roma, Italy
}

\begin{abstract}
Multiphoton indistinguishability is a central resource for quantum enhancement in sensing and computation. Developing and certifying large scale photonic devices requires reliable and accurate characterization of this resource, preferably using methods that are robust against experimental errors. Here, we propose a set of methods for the characterization of multiphoton indistinguishability,  based on measurements of bunching and photon number variance. Our methods are robust in a semi-device independent way, in the sense of being effective even when the interferometers are incorrectly dialled. We demonstrate the effectiveness of this approach using an advanced photonic platform comprising a quantum-dot single-photon source and a universal fully-programmable integrated photonic processor. Our results show the practical usefulness of our methods, providing robust certification tools that can be scaled up to larger systems.
\end{abstract}

\maketitle

\section{Introduction}

Multiparticle interference is one of the fundamental resources for quantum information tasks. In many situations, interference provides a tool to outperform classical strategies for tasks in metrology and information processing. In photonics, multiphoton interference has been exploited for advantage in quantum sensing \cite{giovannetti2011advances,polino2020photonic,albarelli2020perspective}, reaching sensitivities beyond the standard quantum limit. In quantum computation, multiphoton interference enables a non-universal model known as Boson Sampling \cite{aaronson2011computational}, which involves sampling from the output distribution of indistinguishable photons interfering in a linear interferometer.  Boson Sampling machines, featuring variants of the original proposal \cite{hamilton2017gaussian,Lund_SBS,Brod2019_review,Bentivegna15}, have been experimentally implemented  \cite{QAdv1,QAdv2,QAdv3,QAdv4} for demonstrations of quantum computational advantage in Gaussian Boson Sampling experiments that are hard to simulate by classical supercomputers.

Embracing the fundamental role of quantum many-body interference, it is important to develop robust and accurate methods for the characterization of multiphoton indistinguishability. This is essential for the development of better photon sources  \cite{bouchard2020two}, but also for the certification of quantum advantage in applications featuring multiphoton interference. For pairs of photons, the Hong-Ou-Mandel (HOM) interference effect \cite{hom1987} can be used to completely characterize the level of two-particle indistinguishability.  The Hong-Ou-Mandel visibility at the output of a symmetric beam-splitter leads to a direct estimation of the overlap between the two single-photon spectral functions, describing their internal degrees of freedom. If we have three or more photons,   pairwise Hong-Ou-Mandel interference experiments are no longer sufficient to completely characterize indistinguishability \cite{menssen2017distinguishability, shchesnovich18}.

A number of approaches to the characterization of multiphoton indistinguishability have been reported in the literature. One approach has defined inequalities whose violations bound parameters that characterize indistinguishability \cite{Brod19,GalvaoB20,Giordani2020,Giordani2021} or other relevant quantities \cite{wagner2022coherence,giordani2023certification} for multiphoton scenarios. Starting from a prior knowledge of the model describing the multiparticle state, Bayesian \cite{viggianiello2018optimal} or maximum-likelihood \cite{renema2021sample} methods have also been used to perform a direct estimation of the indistinguishability parameters by measuring the output pattern after interference within linear interferometers. The flexibility in the design of linear interferometers requires an identification of the transformations that optimize the amount of information retrieved with respect to data sample size  \cite{viggianiello2018optimal}. Other approaches are based on using specific classes of interferometers. For instance, a generalization of the Hong-Ou-Mandel to symmetric multimode interferometers leads to a class of suppression laws \cite{tichy2010zero,tichy2014stringent,crespi2015supp,Crespi16,dittel2017hypercubes,viggianiello2018experimental}, which identify suppressed events due to the joint effect of interferometer symmetry and bosonic statistics.  Another recent approach uses cyclic interferometers \cite{Pont22} for the estimation of a notion of genuine indistinguishability for multiphoton states of a particular form \cite{Brod19}. 

The methods discussed above are based on measuring output patterns after linear-optics interference and are sensitive to errors in the implementation of each interferometer design. A good characterization of multiphoton indistinguishability requires an accurate characterization of the implemented linear-optical dynamics  \cite{hoch2023characterization}, with characterization errors in principle leading to biases in the analysis of multiphoton indistinguishability. Recent efforts to devise semi-device independent approaches have the goal of enabling accurate conclusions with less need for trusting the characterization of parts of the experiment \cite{vandermeer2021witness}. Here, in Sec. \ref{sec:theory} we propose semi-device independent methods for the characterization of multiphoton indistinguishability, based on measurements of bunching and variance of photon number at the device's output. We then experimentally test the proposed methods in Sec. \ref{sec:exp_verification} using a photonic platform featuring a single-photon quantum dot source and an integrated photonic processors (IPP) that is universal and fully programmable. 

\section{Theory}
\label{sec:theory}

The set-up we consider consists of an $n$-mode general linear-optical interferometer, with $n$ single-photon inputs, one per photonic mode. The interferometer is described by a unitary $n \times n$ matrix $U$, which maps input creation operators to output creation operators. The indistinguishability of the photons is completely described by a Gram matrix of inner products: $S_{i,j}=\left\langle \chi_i |\chi_j \right\rangle$, where $\ket{\chi_i}$ is the spectral function describing all the internal degrees of freedom of the photon entering input port $i$. This Gram matrix can be written only in terms of physically relevant unitary invariant properties of the set of spectral functions describing the single-photon internal degrees of freedom, as described in \cite{Oszmaniec24}.

We will now discuss the theoretical framework that allows for a characterization of multiphoton indistinguishability based on measurements of photonic bunching and photon number variance at the output of a general linear-optical device. 

\subsection{Full bunching probabilities}
The probability of full bunching, studied for example in \cite{tichy2015sampling, spagnolo2013general, aaronson2012generalizing}, is defined as the probability that all $n$ photons leave the interferometer in a single chosen mode. It is known that the probability of full bunching $p_{FB}$ decreases exponentially. More precisely, \cite{aaronson2012generalizing} shows that $p_{FB}$ is upper-bounded by $p_{FB}\le n!/n^n$, with the bound saturated by a balanced interferometer, i.e. one which distributes any single-photon input uniformly among the output modes. While this prevents meaningful measurements of $p_{FB}$ for arbitrarily large numbers of photons, for small photon numbers one can estimate $p_{FB}$ with photon-counting measurements only on a single output mode.

Let us assume one can fix the interferometer design, and pick two different indistinguishability scenarios characterized by Gram matrices $S_1$ and $S_2$. Tichy showed in \cite{tichy2015sampling} that the ratio of $p_{FB}$ for the two scenarios $r_{FB} = p_{FB}(S_1)/p_{FB}(S_2)$ is exactly equal to the ratio of the permanents of the Gram matrices $S_1,S_2$ describing them:
\begin{equation}
\frac{p_{FB}(S_1)}{p_{FB}(S_2)}= \frac{\mathrm{Per}(S_1)}{\mathrm{Per}(S_2)}. \label{eq:fb}
\end{equation}
Let us look at one particular application of Eq. (\ref{eq:fb}) above. A set of $n$ perfectly indistinguishable photons is described by a Gram matrix $S_{1,i,j}=1 \; \forall i,j$, for which $\mathrm{Per}(S_{1})=n!$. On the other hand, if the photons are perfectly distinguishable, they will be described by a diagonal Gram matrix $S_{2,i,j}=\mathrm{diag}(1,1,\dots,1)$, for which $\mathrm{Per}(S_{2})=1$, giving a ratio of full bunching for these two scenarios that is $\mathrm{Per}(S_1)/\mathrm{Per}(S_2)=n!$. This is the full bunching law of \cite{spagnolo2013general}, which Eq. (\ref{eq:fb}) generalizes for arbitrary indistinguishability scenarios featuring pure single-photon spectral functions.

The law in Eq. (\ref{eq:fb}) is valid for any pair of Gram matrices. As we will see in the section on experimental results, this means one can fix an \textit{arbitrary} interferometer design, then measure the probability of full bunching for perfectly distinguishable photons [described by a Gram matrix $S_2=\mathrm{diag}(1,1,\dots,1)$]. We can then prepare inputs described by a different Gram matrix $S_1$, and again measure $p_{FB}$. According to Eq. (\ref{eq:fb}), the ratio of measured full-bunching probabilities in the two scenarios will give us an estimate for $\mathrm{Per}(S_1)/\mathrm{Per}(S_2)=\mathrm{Per}(S_1)$, which can be seen as a measure of photon indistinguishability \cite{tichy2015sampling}.
Another point worth noting is that $p_{FB}$ is a function of $\mathrm{Per}(S)$, which itself in general depends on higher-order unitary invariants encoded in the Gram matrix (see \cite{Oszmaniec24} for an overview). This means that measurements of $p_{FB}$ are sensitive to these so-called collective photonic phases \cite{menssen2017distinguishability, shchesnovich18}, and can be used to estimate them, as we will see in the section on experimental results.

\subsection{Photon number variance}
We will now show that the average photon number variance at the output of the device can also be used to characterize multiphoton indistinguishability. We define this quantity as:
\begin{equation}
\label{eq:variance_definition}
\sigma=\frac{1}{n}\sum_{i=1}^n \left\langle n_i^2 \right\rangle - \left\langle n_i \right\rangle^2 ,
\end{equation}
where $n_i$ is the number of photons at output mode $i$, still assuming we inject exactly one photon per input mode. Unlike the probability of full bunching, $\sigma$ does not decrease as $n$ increases. On the other hand, its estimation is experimentally challenging, as in principle it requires photon-counting detection on all output modes.

In App. \ref{ap:sigma} we obtain a universal and simple expression for $\sigma$ that is valid for any interferometer and any number of modes:
\begin{equation}
\sigma= 1 + \frac{1}{n}\sum_{a \neq b} \sum_i |\langle \chi_a | \chi_b \rangle|^2 |U_{ia}|^2 |U_{ib}|^2 - \frac{1}{n}\sum_{ik} |U_{ik}|^4.  
\label{eq:sigma_U}
\end{equation}
Note that, unlike the probability of full bunching $p_{FB}$, the average photon number variance $\sigma$ only depends on two-photon overlaps $\Delta_{ab}=|\langle \chi_a | \chi_b \rangle|^2$, and not on higher-order unitary invariants written as functions of three or more spectral functions \cite{Oszmaniec24, menssen2017distinguishability, shchesnovich18}. Moreover, it depends only on the absolute value squared of the matrix elements of the interferometers, which can be measured using classical light.

It can be seen from Eq. (\ref{eq:sigma_U}) that any increase on the overlaps $\Delta_{ab}$ will result in an increased variance. This implies that, for a given interferometer, the variance is maximized when photons are fully indistinguishable. We show in App. \ref{ap:sigma} that the overall maximum of the variance is obtained by further optimizing over possible interferometers, which leads to the value
\begin{equation}\label{eq:sigma_max}
    \sigma^{\max}=2-2/n,  
\end{equation}
obtained for indistinguishable photons and balanced interferometers on $n$ modes, i.e. where $|U_{ij}|^2=1/n$ such as the Fourier interferometer. In fact, it can be seen that the implementation of a balanced interferometer gives us direct access to the average indistinguishability of the input states since 
\begin{equation}\label{eq:sigma_bal}
    \sigma^{bal}= 1+ \frac{n-1}{n}\bar{\Delta}-\frac{1}{n}.
\end{equation}
Here, we defined the average indistinguishability $\bar{\Delta}$ as
\begin{equation}\label{eq:av_Delta}
   \bar{\Delta}=\frac{\sum_{a\neq b} \Delta_{ab}}{n(n-1)}. 
\end{equation}
For large $n$, the expression in Eq.~\eqref{eq:sigma_bal} matches the one obtained in Ref. \cite{robbio2024central} by different techniques, which use the fact that the reduced density matrix of a single output of the interferometer is well described by a product of thermal states in the asymptotic limit.

We also show in App. \ref{ap:bound_Delta_min} that the measurement of the variance with the Fourier interferometer for $n=3$  also gives us a non-trivial lower bound for all two-photon overlaps $\Delta_{ab}$: 
\begin{equation}
\label{eq:min_delta_tritter}
\min_{a\neq b} \Delta_{ab} \ge \left(\frac{9}{4}\sigma-2 \right)^2.
\end{equation}
without the need for any assumptions on the form of the Gram matrix $S$. This shows $\sigma$ can be used to certify the minimum value of all 3 two-photon indistinguishabilities.

While balanced interferometers give us a direct way to obtain guarantees on indistinguishability, we can also show that \emph{arbitrary} interferometers can be used to reconstruct the matrix of overlaps $\Delta_{ab}$. Assuming knowledge on the values of $|U_{ij}|^2$, the measurement of $\sigma$ for different interferometers allows us to reconstruct the matrix $\Delta$ by solving a linear system of equations (see Sec.~\ref{sec:exp_variance}). This approach is a practical way of extracting information on the indistinguishability of the input photons directly from the photon number variance $\sigma$ at the output of arbitrary interferometers. This could be helpful in monitoring the photon source indistinguishability with no need for interrupting the data acquisition to perform Hong-Ou-Mandel tests.  

We may also assume a scenario where we do not trust the implementation of some desired unitary matrices, either because of experimental imperfections, or because we have access to an interferometer built by an untrusted party. Even then it may be possible to use such a device to obtain guarantees on the average indistinguishability $\bar{\Delta}$. This can be done by measuring the value of $\sigma$ for a given distinguishability scenario defined by the matrix $\Delta$ as well as the value of $\sigma^d$, i.e. the average variance when distinguishable photons are sent to the same device. In App. \ref{ap:bounds_average_delta}, we prove the following inequality, which is valid for any interferometer 
\begin{equation}\label{eq:bound_avDelta}
    \bar{\Delta}\geq \frac{1}{n(n-1)}\left(\frac{\sigma-2\sigma^d+1}{1-\sigma^d}\right)^2-\frac{1}{n-1} = \bar{\Delta}_{LB}.
\end{equation}
This bound can be used to give us a guarantee on photon indistinguishability in a semi-device independent scenario, where the preparation of single photons and the Fock measurements are trusted, but not the interferometer.

\section{Experimental verification}
\label{sec:exp_verification}
As a next step, we performed experimental measurements of the probability of full bunching $p_{FB}$ and the average photon number variance at the output $\sigma$. The employed platform comprises an up-to-date solid-state single-photon source, with the aim of characterizing the indistinguishability of the output photons and of testing different regimes via controlled tuning. We implemented specific interferometers, such as the balanced tritter, as well as randomly chosen $3 \times 3$ transformations, as a way to test the semi-device independence features of our approach. In Sec. \ref{sec:exp_setup} we describe the experimental setup used for our multiphoton indistinguishability tests. In Sec. \ref{sec:exp_full_bunching_ratio} we describe the results on the full bunching ratio measurements, and in Sec. \ref{sec:exp_variance} we discuss the experimental results related to the measurement of the average photon number variance.

\subsection{Experimental setup}
\label{sec:exp_setup}

The experimental apparatus corresponds to a photonic machine named QOLOSSUS, that comprises different components used for photon generation, manipulation and detection. 
\begin{figure*}
    \centering
    \includegraphics[width=0.99\linewidth]{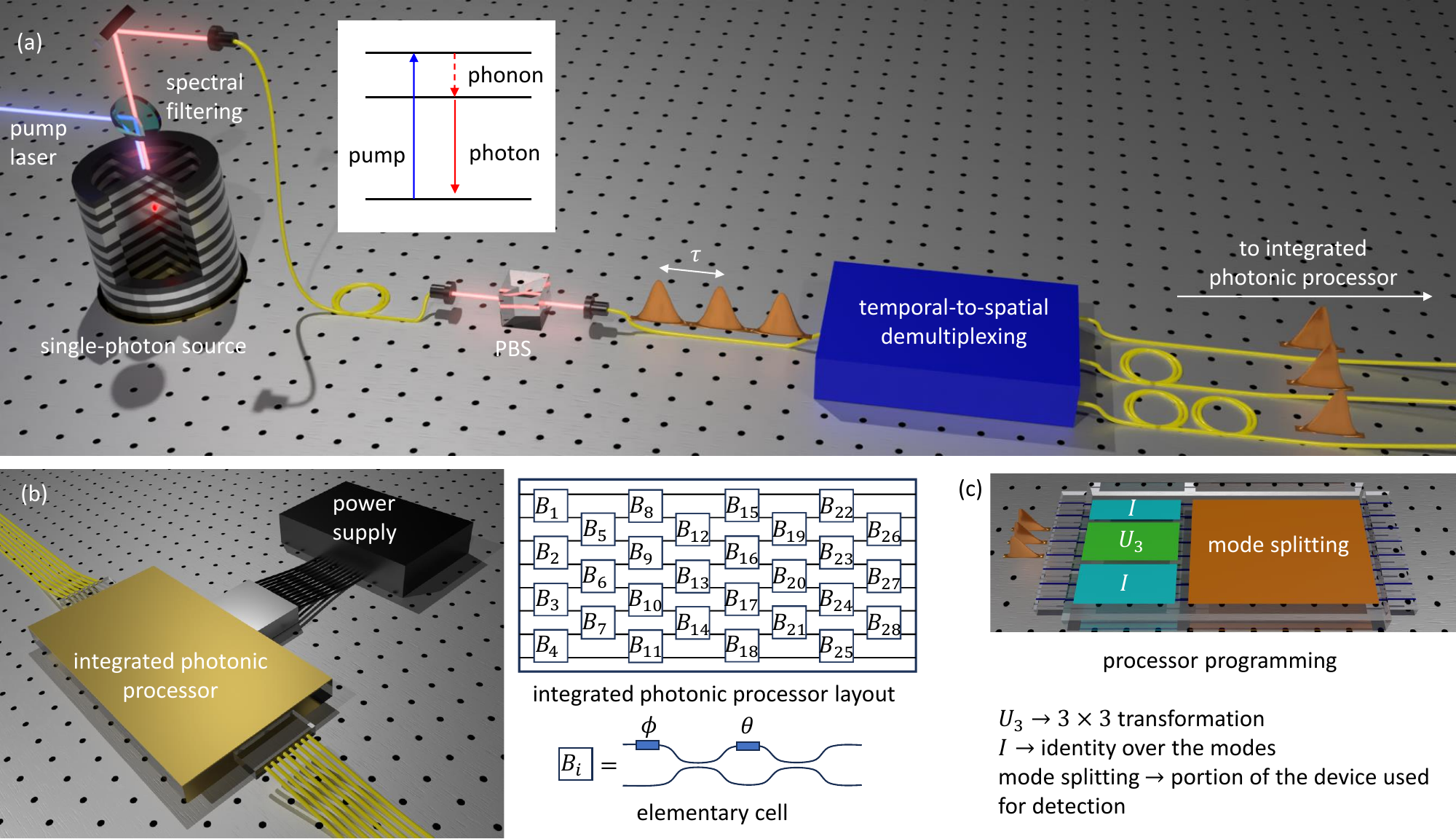}
    \caption{\textbf{Experimental setup.} QOLOSSUS machine employed for the verification of the proposed tests. (a) Single-photon source composed by a quantum dot operating in a non-resonant regime, emitting a train of single-photon pulses at fixed time intervals $\tau$. A demultiplexing module converts such a sequence to a set of three-photon input states injected in different modes of an IPP. (b) Scheme of the universal 8-mode integrated photonic chip, using the layout of \cite{Clements16}. The elementary cells $B_i$ are programmable MZIs composed of two balanced directional couplers and of two thermo-optic phase shifters. Photons are coupled at the input and at the output of the integrated processor via single-mode fiber arrays, directly pigtailed to the device. (c) Device programming scheme for the different performed measurements. When the processor is employed to measure the full bunching ratio (top), three photons are injected in modes [3,4,5], with the first three layers of elementary cells used to program transformation $U_3$. The subsequent layers, and an off-chip fiber beam-splitter, are used to perform pseudo photon number resolving detection via avalanche photo-diodes and a time-tagging system. $I$ corresponds to the identity over the modes.}
    \label{fig:experimental_setup}
\end{figure*}
The multi-photon input states are generated using a quantum-dot single-photon source (Quandela \textit{e-Delight} system), consisting of an InGaAs matrix located in a nanoscale electrically controlled micropillar cavity \cite{gazzano2013bright, nowak2014deterministic, somaschi2016near, ollivier2020reproducibility, thomas2021bright}. The source operates in a cryogenic environment (approximately 4K), obtained within a \textit{Attocube-AttoDry800} He closed-cycle cryostat, and works in a non-resonant regime exploiting a longitudinal-acoustic (LA) phonon-assisted configuration \cite{thomas2021bright}, pumped by a blue-detuned excitation obtained with a 927.2 mm wavelength, $R_{\text{exc}} = 79$ MHz repetition rate, pulsed pump laser. The photons are emitted with a wavelength of 927.8 nm, in a train of pulses emitted in time-bin of fixed distance $\tau \sim 12.5$ ns, that is, the inverse of the pulsed pump laser repetition rate. The output photons are coupled in a single-mode fiber through a confocal microscope mounted on top of the cryostat and then separated from the residual laser pump via spectral filtering. Here we measure, on avalanche photodiodes with a quantum efficiency of $\eta_{\text{d}} \sim 35\%$, a typical count rate of $R \sim 3.5$ MHz. Accounting for the transmission efficiency of the collection setup, estimated to be $\eta_{\text{c}} \sim 55 \%$, this corresponds to a first lens brightness $B = R/(R_{\text{exc}}\eta_{\text{d}}\eta_{\text{c}}) \sim 23\%$. We assessed the single photon purity and pairwise photon indistinguishability by measuring the second-order autocorrelation function and the visibility in a Hong-Ou-Mandel experiment, obtaining $g^{(2)}(0) \sim 0.02$ and $V^{\mathrm{HOM}}\sim 0.91$, respectively. The train of single-photon pulses emitted by the source is then converted in a multi-photon input state, corresponding to single photons propagating in different spatial modes, via a commercially available temporal-to-spatial demultiplexing system \textit{Quandela QDMX-6} \cite{pont2022high,Pont22}. Specifically, the system employs a RF-modulated acousto-optical material to diffract an input signal towards several output spatial modes. We optimized the demultiplexing system to continuously divide the input signal into three bunches with a duration of $T_b \sim 180$ ns each. To achieve temporal synchronization of the photon bunches, we added time delays obtained via custom-length single-mode fibers. The overall input-output transmission efficiency has been measured to be in the interval $\eta_{\text{DMX}} \sim 0.75$ in the worst case. The typical degree of HOM visibility $V(T)$ among photons separated by a time interval $T$ was found to be $V^{\mathrm{HOM}}_{12(23)}(180 \; \text{ns}) = 0.83(1)$ and $V^{\mathrm{HOM}}_{13}(360 \; \text{ns}) \sim 0.81(1)$. This visibility decreases with increased pairwise photon delays, as already observed in \cite{pont2022high}. The reason for the visibility decrease could be related to both the acousto-optic effect at the basis of the demultiplexing system, and to spectral wandering effects of the QD emission on the timescale of several hundreds of nanoseconds. The overall source configuration is shown in Fig. \ref{fig:experimental_setup}(a).

We use an 8-mode, fully programmable IPP to characterize the multiphoton indistinguishability of the source. This device, fabricated via the femtosecond laser writing (FLW) technology \cite{corrielli2021flm}, consists of a mesh of waveguide interferometers arranged according to the rectangular layout of \cite{Clements16}. This layout choice allows the implementation of any linear-optical transformation on 8 modes. As shown in Fig. \ref{fig:experimental_setup}(b), the device is composed of a network of 28 beam-splitters with variable beam-splitting ratios acting as elementary cells $B_i$, each of which actually implemented as programmable Mach-Zehnder Interferometers (MZIs). Each MZI is implemented using two cascaded balanced directional couplers, and two thermo-optic phase shifters which provide full control of the cell operation. The fabrication and calibration procedures are detailed in App. \ref{ap:processor}.

\subsection{Measuring the full bunching ratio}
\label{sec:exp_full_bunching_ratio}

As a first step, we programmed the device to implement a balanced tritter transformation $U_{3}$ on three single-photon inputs, injected in modes [3,4,5] [see Fig. \ref{fig:experimental_setup}(c)]. The device's programming was performed in the following way: the first three layers of the IPP were used to implement the linear-optical unitary $U_3$, while the remainder of the device, together with an external in-fiber beam splitter, was used to split the 3 outputs of $U_3$ to perform pseudo photon number resolving detection. The implemented transformation $\tilde{U}_3$ has been found to reach a fidelity of $F = \vert \mathrm{Tr}(U_3 \tilde{U}_3^{\dag}) \vert/3 > 0.999$ w.r.t. the ideal unitary $U_3$ (see App. \ref{ap:interferometer}). Hereafter, we will label as $\sim$ all the experimentally measured quantities. The general form of a Gram matrix describing a three-photon indistinguishability scenario is given by a matrix in the form:
\begin{equation}
\label{eq:Sk}
S_k = 
\begin{pmatrix}
1 & \sqrt{\Delta^{k}_{ab}} & \sqrt{\Delta^{k}_{ac}} \\
\sqrt{\Delta^{k}_{ab}} & 1 & \sqrt{\Delta^{k}_{bc}} e^{i\varphi}\\
\sqrt{\Delta^{k}_{ac}} & \sqrt{\Delta^{k}_{bc}} e^{-i\varphi}& 1
\end{pmatrix}.
\end{equation}
In what follows, we parameterize a scenario's indistinguishability using positive, real-valued Gram matrices $S_k$, i.e. we assume $\varphi = 0$, while $k$ labels the scenario and $(a,b,c)$ labels the different photons injected at the interferometer's inputs. The assumption that the indistinguishability is well-described by a Gram matrix as above was verified experimentally via two independent tests, that is, by using a methodology based on outcomes of a cyclic interferometer \cite{Pont22}, and by direct analysis of the output distribution after the tritter (see App. \ref{ap:cyclic}). At first, we implemented a scenario described by a Gram matrix $S_A$ that identifies the intrinsic photon source indistinguishability reached after the demultiplexing module. In this configuration, the values $\tilde{\Delta}^{A}_{ij}$ are estimated from the visibilities $V^{\mathrm{HOM}}_{ij}$ of independent Hong-Ou-Mandel experiments, by correcting for the effect of the non-zero value of the second order correlation function $g^{(2)}(0) = 0.0218(6)$ following the approach of \cite{Olli21}: $\tilde{\Delta}_{ij} = (V^{\mathrm{HOM}}_{ij} + g^{(2)}(0))/(1 - g^{(2)}(0))$. In our case, we obtained $\tilde{\Delta}^{A}_{ab} = 0.875(4)$, $\tilde{\Delta}^{A}_{ac} = 0.874(2)$ and $\tilde{\Delta}^{A}_{bc} = 0.848(2)$.

The second scenario corresponds to a diagonal Gram matrix $S_D$ describing photons made completely distinguishable due to other auxiliary degrees of freedom, that is, time and polarization. Tuning indistinguishability using polarization is made possible by the polarization-independent operation of the IPP.

\begin{figure}[ht!]
\centering
\includegraphics[width=0.49\textwidth]{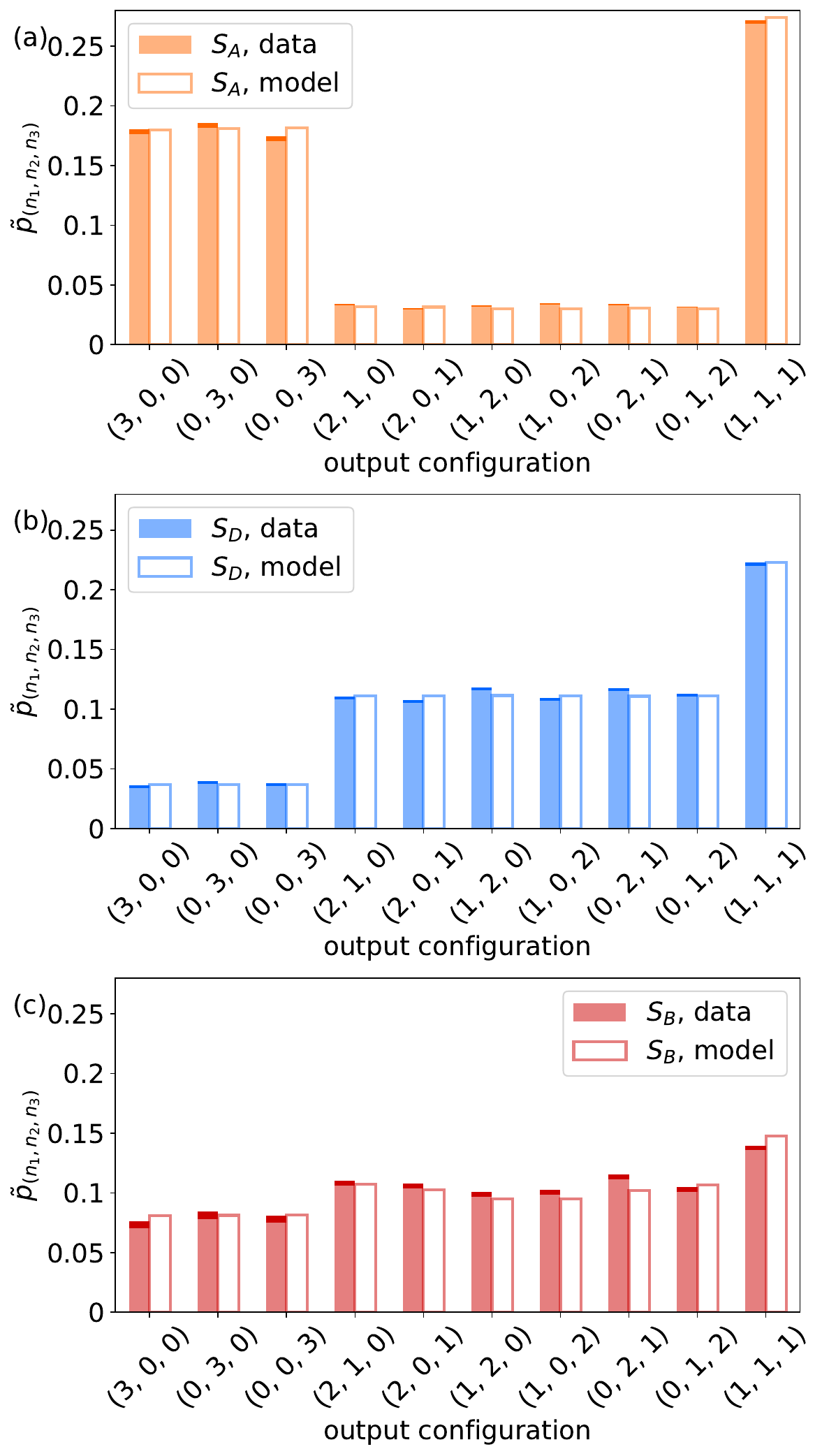}
\caption{\textbf{Experimental measurement of the output distributions with a balanced tritter.} The measured distributions $\tilde{p}_{(n_1, n_2, n_3)}$ are obtained by programming the device to act as the $\tilde{U}_{3}$ transformation according to the discussion in the main text. (a) Distribution obtained for a Gram matrix $S_A$. (b) Distribution obtained for distinguishable particles, corresponding to a Gram matrix $S_D$. (c) Distribution obtained for a Gram matrix $S_B$. Colored bars: experimental data, darker regions correspond to 1-$\sigma$ confidence intervals. White bars: predictions obtained from a model taking into account the implemented transformation $\tilde{U}_{3}$, the Gram matrices estimated from pairwise Hong-Ou-Mandel experiments, losses, and multiphoton contributions obtained from the estimated $g^{(2)}$ value.}
\label{fig:data_distr_tritter}
\end{figure}

The measured output probability distributions with 3-photon inputs described by those two indistinguishability scenarios are shown in Fig. \ref{fig:data_distr_tritter} (a-b), and compared with a model (see App. \ref{ap:model}) which includes all main experimental features such as losses, the effective matrix $\tilde{U}_{3}$, the effective Gram-matrix of the input configuration, and multiphoton contributions related to the non-zero $g^{(2)}(0)$. 
The data are in good agreement with the expectation from the model, see App. \ref{ap:data_analysis} for more details on the analysis. The agreement can be quantified by the total variation distances $\mathrm{TVD} = 1/2 \sum_i \vert p_i - q_i \vert$ between experiment ($p_i$) and model ($q_i$). The obtained values for the two configurations are respectively $\mathrm{TVD}(S_{A}) = 0.015(2)$ and $\mathrm{TVD}(S_{D}) = 0.013(2)$. 

\begin{figure*}[ht!]
\centering
\includegraphics[width=0.99\textwidth]{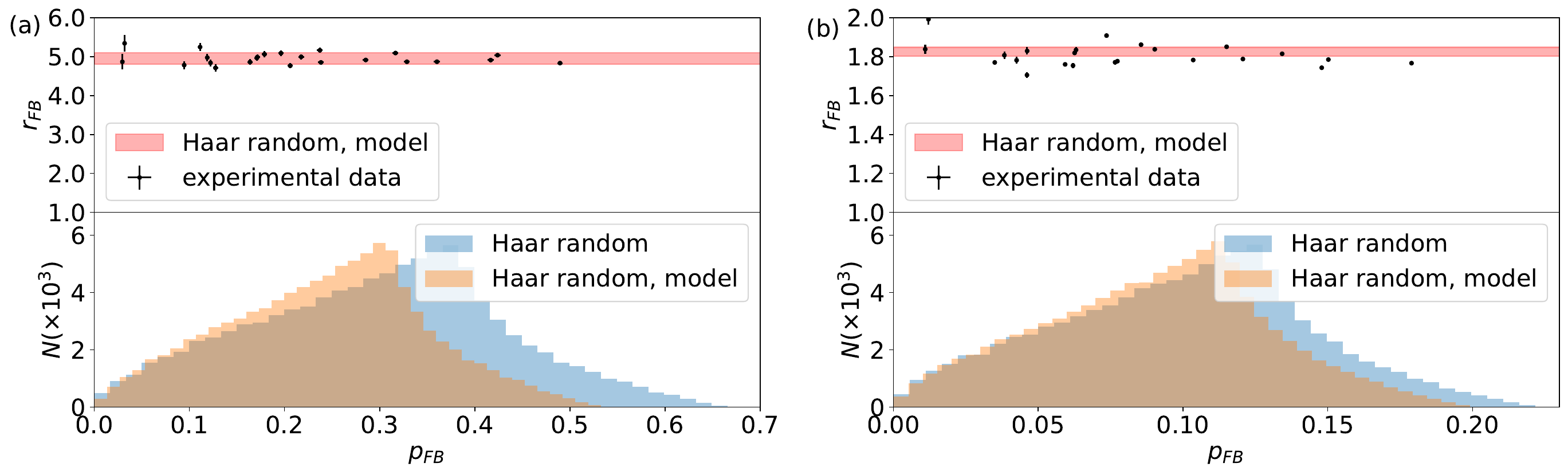}
\caption{\textbf{Measurement of the full bunching ratios for 23 randomly-drawn interferometers.} Measured values of $r_{FB}$ in the scenario $S_{A}$ (a) and in the scenario $S_{B'}$ (b) as a function of the corresponding measured $p_{FB}$. The red shaded area in the two scenarios corresponds to a prediction obtained by averaging the expected value with the model taking into account losses and multiphoton emission from the source, and fluctuation on pairwise indistinguishability during the experimental run. The bottom histograms show the distribution of $p_{FB}$ over Haar random unitaries, obtained via a Monte Carlo simulation in the ideal scenario in the two indistinguishability configurations (blue) and taking into account experimental imperfections (orange).}
\label{fig:data_haar_3}
\end{figure*}

From the measured samples we have then estimated the full bunching ratios for the two scenarios $[\tilde{p}_{FB}(S_A), \tilde{p}_{FB}(S_D)]$, summed over all the three possible output configurations $[(3,0,0), (0,3,0), (0,0,3)]$. The measured value is found to be $\tilde{r}_{FB}(S_A) = \tilde{p}_{FB}(S_A)/\tilde{p}_{FB}(S_D) = 4.9(1)$. This is to be compared in principle with the ratio between the permanents of the Gram matrices $\mathrm{per}(S_A)/\mathrm{per}(S_D) = 5.21(1)$. A more accurate prediction of the full bunching ratio value expected from the experiment can be computed from the model that takes into account all the other sources of noise in the setup, i.e. multiphoton contributions and optical losses as described in App. \ref{ap:model}. More specifically, by including these effects the expected value reduces from the ratio between the Gram matrix permanents $\mathrm{per}(S_A)/\mathrm{per}(S_D) = 5.21(1)$ to $r_{FB}(S_A) = 4.91(1)$, thus showing a very good agreement with the experimental data. We use Eq. (\ref{eq:fb}) and the two output distributions to estimate the full bunching ratio for each of the 3 output modes of $U_3$: $\{\tilde{r}^{(3,0,0)}_{FB}(S_A), \tilde{r}^{(0,3,0)}_{FB}(S_A), \tilde{r}^{(0,0,3)}_{FB}(S_A)\} = \{5.1(2), 4.8(2), 4.7(2)\}$. The obtained values are in good agreement with the predictions from the model, which are found to be $\{r^{(3,0,0)}_{FB}(S_A), r^{(0,3,0)}_{FB}(S_A), r^{(0,0,3)}_{FB}(S_A)\} = \{4.91(1), 4.91(1), 4.91(1)\}$. 

As a further test of our approach, we prepared photons described by a different Gram matrix $S_B$ by tuning the polarization state of photon $b$ to reduce its indistinguishability with respect to the other two. The corresponding overlaps estimated from the HOM visibilities were $\tilde{\Delta}_{ab} = 0.103(4)$, $\tilde{\Delta}_{ac} = 0.881(2)$, $\tilde{\Delta}_{bc} = 0.107(2)$. The measured distribution is in good agreement with the predictions from the model [see Fig. \ref{fig:data_distr_tritter} (c)], quantified by the total variation distance $\mathrm{TVD}(S_B) = 0.025(4)$. This configuration led to a measured full bunching ratio of $\tilde{r}_{FB}(S_B) = \tilde{p}_{FB}(S_B)/\tilde{p}_{FB}(S_D) = 2.11(6)$, to be compared with the ratio between the permanents of the Gram matrices $\mathrm{per}(S_B)/\mathrm{per}(S_D) = 2.28(2)$ and with the prediction obtained from the model $r_{FB}(S_B) = 2.21(2)$. 

To experimentally show that our approach is robustly independent of the interferometer implemented, we performed additional experiments where the IPP was programmed to implement $23$ different random $3 \times 3$ trasformations, obtained by driving with random currents the thermo-optic phase shifters of the $U_3$ block of the IPP [see Fig. \ref{fig:experimental_setup}(c)]. We measured the full bunching ratio $r_{FB}$ of two indistinguishability scenarios against the one with distinguishable photons. Again, scenario $S_{A}$ features highly indistinguishable photons from the source after demultiplexing. Conversely, we reconstruct a scenario $S_{B'}$ in which one photon is completely distinguishable from the other two, corresponding to overlaps $\tilde{\Delta}^{B'}_{ab} = 0$, $\tilde{\Delta}^{B'}_{ac} = 0.831(4)$, $\tilde{\Delta}^{B'}_{bc} = 0$, by injecting two photons in the interferometer inputs $\{a,c\}$ and reconstructing the full bunching probabilities from two-fold coincidence counts (see App. \ref{ap:data_analysis}). The results are shown in Fig. \ref{fig:data_haar_3} as a function of the full bunching probability $p_{FB}$ for each random unitary. We observe that in both cases the measured values of $r_{FB}$ are very close to the expected ones, independently of the dialled transformations. 

\subsection{Measuring the photon number variance}
\label{sec:exp_variance}
The second step was dedicated to the characterization of photon indistinguishability based on measurement of the photon number variance in our experimental platform. This approach requires measurement of the probabilities for all possible photon number output configurations, done via the detection system described above, based on mode splitting and coincidence detection. 

We considered the same indistinguishability scenario $S_A$ and $S_B$ described above, and evaluated the variance $\tilde{\sigma}(S_k)$ using the same measured data sample corresponding to a device programmed to act as a balanced tritter. The estimated value of the variance was measured to be $\tilde{\sigma}(S_A) = 1.199(4)$, while the maximum value for fully indistinguishable photons would be $4/3$ [see Eq.~\eqref{eq:sigma_max}]. The bound presented in Eq. (\ref{eq:min_delta_tritter}) can then be used to determine a non-trivial lower bound $\mathrm{min} \Delta_{ij} \geq 0.49(1)$ for the three individual overlaps. The same analysis was performed on data collected using the indistinguishability scenario $S_{B}$, resulting in an estimated value of the variance $\tilde{\sigma}(S_{B}) = 0.885(3)$. As expected, the previous analysis now fails to give a non-trivial lower bound for the overlaps values, being below the minimum threshold $8/9$ for Eq. \eqref{eq:min_delta_tritter} to give a meaningful lower bound on $\min \Delta_{ij}$.

Similarly, since the interferometer $U_3$ was set to implement the tritter matrix, we can use the relationship given by Eq. \eqref{eq:sigma_bal} to extract an estimation of the average pairwise indistinguishability in the two scenarios given by:
\begin{equation}
    \begin{split}
        \bar{\Delta}(S_A) = 0.80(1), \\
        \bar{\Delta}(S_B) = 0.33(1).
    \end{split}
\end{equation}
These values are found to be compatible with the pairwise Hong-Ou-Mandel visibilities for each photon pair. Indeed, the experimental estimate of the photon number variance is sensitive to the presence of multiphoton components in the emitted light, and thus the bound is expected to be compatible with the Hong-Ou-Mandel visibilities, that include both the overal and the multiphoton contributions. Finally by considering the expected variance in the distinguishable scenario for the balanced three-mode interferometer $\sigma^d = 2/3$, we can compute a non-trivial lower bound to the average pairwise indistinguishability $\bar{\Delta}_{LB}$, as in Eq. \eqref{eq:bound_avDelta}:
\begin{equation}
    \bar{\Delta}(S_A) \geq 0.63(1),
\end{equation}
while scenario $S_B$ fails also in this case to give a non-trivial lower bound for the overlaps values. Again, we note that such estimation of a lower bound on $\bar{\Delta}$ is completely semi-device independent, as it requires as a further element only a direct measurement of the average variance in a fully-distinguishable scenario $\sigma^d$ or its reconstruction, as in Eq. \eqref{eq:sigma_U}, via the unitary moduli $\vert U_{ij} \vert$, which can be measured in a straightforward experiment using classical light.

Besides this direct application of the photon number variance $\sigma$ either to obtain a lower bound on all overlaps using Eq. (\ref{eq:min_delta_tritter}) to estimate an average pairwise overlap parameter, we now show how $\sigma$ measurements can also be used to directly monitor the value of each individual overlap. 
\begin{figure}[ht!]
    \centering
    \includegraphics[width=0.46\textwidth]{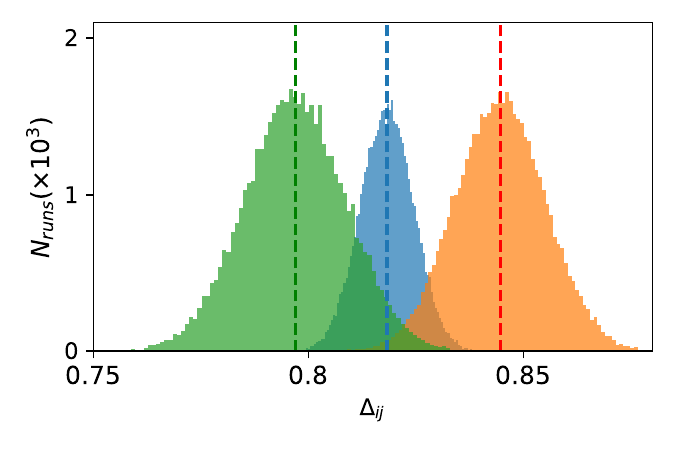}
    \caption{\textbf{Inference of the experimental overlaps (visibilities) via the photon number variances measured over 23 randomly sampled unitaries.} Here we report the histogram of pairwise overlaps $\{\Delta'_{ab},\Delta'_{bc},\Delta'_{ac}\}$ inferred by numerically minimizing Eq. \eqref{eqn:optvar} over 50000 bootstrapped optimization runs. The considered photon number variances in the optimization problem are obtained experimentally considering three photons injected in the interferometer set to implement 23 different unitaries, in the indistinguishability scenario $S_A$. Blue bars: $\Delta'_{ab}$. Orange bars: $\Delta'_{ab}$. Green bars: $\Delta'_{bc}$. Vertical dashed lines identify the average of the Monte Carlo distributions.}
    \label{fig:optsigma}
\end{figure}
The key idea is to program the device to implement interferometers drawn from a random ensemble, and use $\sigma$ together with information from measurements of the moduli $\vert U_{ij} \vert$ as we now describe. We can interpret Eq. (\ref{eq:sigma_U}) as a linear equation for the three unknown overlaps $\vert \langle \chi_a \vert \chi_b \rangle \vert^2$, $\vert \langle \chi_b \vert \chi_c \rangle \vert^2$ and $\vert \langle \chi_c \vert \chi_a \rangle \vert^2$ we would like to estimate, while the moduli $\vert U_{ij}\vert$ and $\sigma$ as measured (known) coefficients. If we dial 3 different unitaries and measure the corresponding variances $\sigma$, together with knowledge of the moduli $\vert U_{ij} \vert$, Eq. (\ref{eq:sigma_U}) gives us a linear system of equations that can be solved to obtain the three individual overlap values. The robustness of the method to experimental imperfections, e.g., small changes in the overlaps during the measurement runs, can be improved by collecting an overcomplete data set corresponding to more than 3 unitaries, and performing a numerical optimization for a more stable solution. One way of doing this is to consider a given set of unitaries $\{ U_{i} \}$, and define the following numerical minimization problem:
\begin{equation}
    \min_{\Delta_{jk}} \sum_i \frac{[\tilde{\sigma}_{U_i} - \sigma_{U_i}(\Delta'_{ab},\Delta'_{ac},\Delta'_{bc})]^2}{\text{Var}[\tilde{\sigma}_{U_i}]},
    \label{eqn:optvar}
\end{equation}
where $\text{Var}[\tilde{\sigma}_{U_i}]$ is the experimental uncertainty associated with the photon number variance $\tilde{\sigma}_{U_i}$ measured when the interferometer is programmed to implement unitary $U_i$.  
\begin{figure}[b!]
    \centering
    \includegraphics[width=0.99\linewidth]{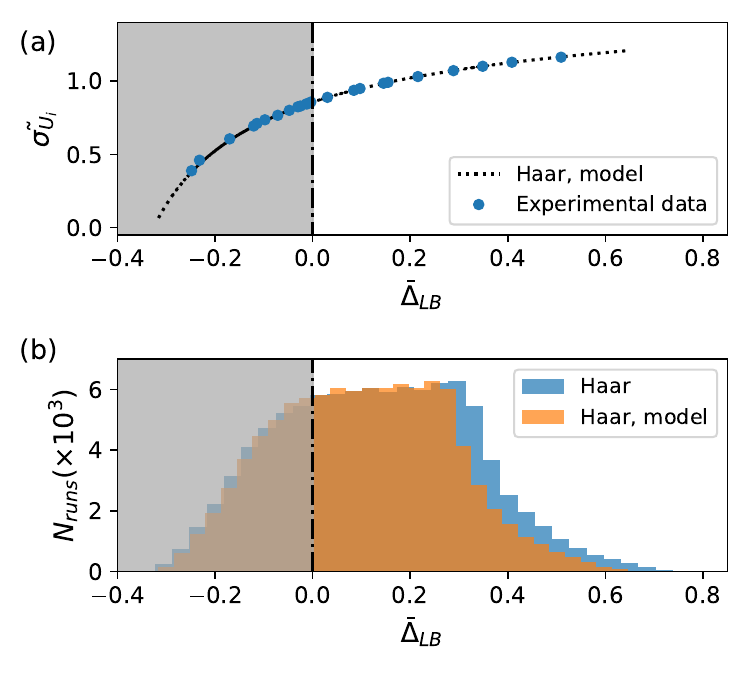}
    \caption{\textbf{Semi-device independent lower bounds on the average overlap $\bar{\Delta}$:} (a) Lower bounds $\bar{\Delta}_{LB}$ on the average overlap parameter $\bar{\Delta}$, inferred from the experimentally measured photon number variances $\tilde{\sigma}_{U_i}$. Here we show $\tilde{\sigma}_{U_i}$ as a function of $\bar{\Delta}_{LB}$ which has been inferred from the data. The dashed line shows the numerically expected behaviour in the indistinguishability scenario described by $S_A$. (b) Histogram of a Monte Carlo simulation of the lower bound on the average overlap parameter $\bar{\Delta}$ obtained from numerical calculation of the photon number variances both in an ideal scenario described by the Gram matrix $S_A$, and taking into account experimental imperfections such as multi-photon contributions - in blue and orange respectively. This gives us numerical evidence that with Gram matrix $S_A$, around $70 \%$ of randomly sampled Haar random unitaries give a non-trivial, i.e. positive, lower bound on the parameter $\bar{\Delta}$.
    On both plots, this is highlighted as the non-shadea area corresponding to non-trivial lower bounds $\bar{\Delta}_{LB}$.}
    \label{fig:lambda_bnd}
\end{figure}
We applied this method to the output data from the 23 implemented random unitaries, in scenario $S_A$. As reported in Fig. \ref{fig:optsigma}, we show the numerical results of 50000 bootstrapped optimizations runs of the quantity in Eq. \eqref{eqn:optvar}. The resulting values for the estimated overlaps $\Delta'_{ab} = 0.818(6)$, $\Delta'_{bc}=0.80(1)$ and $\Delta'_{ac}=0.84(1)$ are found to be compatible with the average Hong-Ou-Mandel visibilities for each photon pair monitored independently during the data acquisition process for the random interferometers. Indeed Eq. (\ref{eq:sigma_U}) does not take into account multiphoton emission between the source, while the presence of multiphoton emission has the effect of reducing the observed value of the Hong-Ou-Mandel visibilities with respect to the overlaps $\vert \langle \chi_{i} \vert \chi_{j} \rangle \vert^2$. To conclude, we note that the semi-device independent bound described by Eq. \eqref{eq:bound_avDelta} returns a positive and thus non-trivial lower bound for the average indistinguishability $\bar{\Delta}_{LB} \geq 0$ for approximately half the implemented random unitaries, as reported in Fig. \ref{fig:lambda_bnd}. There, we also show the distribution of the lower bounds on the average overlap parameter $\bar{\Delta}_{LB}$ obtained via a Monte Carlo numerical simulation over $10^5$ randomly drawn Haar random matrices, carried out in an ideal scenario described by the indistinguishability configuration $S_A$ and finally taking into account experimental imperfections such as multi-photon contributions. This gives us numerical evidence of the fact that for this indistinguishability scenario, around $70\%$ of Haar random unitaries give a non-trivial, i.e. positive, lower bound on the average overlap parameter.

\section{Conclusions}

We have reported on the proposal and experimental implementation of tests to perform robust characterization of multiphoton interference. Our approach, based on measuring the bunching properties or the average  variance of the photon number distribution at each output, can be considered semi-device independent as the results are not biased by incorrect implementation of different interferometer designs. The validity of the approach has been verified experimentally in an advanced photonic platform comprising different components such as a quantum-dot source interfaced with a fully-programmable IPP. In particular, our approach based on photon number variances improves upon previous work on semi-device independent witnesses of indistinguishability based on a single two-mode correlator \cite{vandermeer2021witness}, by giving non-trivial guarantees on indistinguishability for a large fraction of Haar random interferometers. The obtained results have confirmed the robustness of the method, showing its applicability in a practical setting that can be extended  for the characterization of sources of larger numbers of photons. 

\section*{Acknowledgments}
This work was supported by the FET project PHOQUSING (“PHOtonic Quantum SamplING machine” - Grant Agreement No. 899544) and by ICSC – Centro Nazionale di Ricerca in High Performance Computing, Big Data and Quantum Computing, funded by European Union – NextGenerationEU. LN and EFG acknowledge support from FCT -- Fundação para a Ciência e a Tecnologia (Portugal) via project CEECINST/00062/2018. The fabrication of the IPP was partially performed at PoliFAB, the micro- and nanofabrication facility of Politecnico di Milano (www.polifab.polimi.it). The authors would like to thank the PoliFAB staff for the valuable technical support.

\appendix
\section{Average photon number variance at the output} 
\label{ap:sigma}

In this Section we obtain a simple and universal expression for the average photon number variance $\sigma$ that is valid for any interferometer, and any number $n$ of modes: 
\begin{equation}
\sigma= 1 + \frac{1}{n}\sum_{a \neq b} \sum_i |\langle \chi_a | \chi_b \rangle|^2 |U_{ia}|^2 |U_{ib}|^2 - \frac{1}{n}\sum_{ik} |U_{ik}|^4.
\label{ap:exp_var}
\end{equation}
To derive this expression, we will use the fact that correlation between photon numbers in modes $i$ and $j$, with $i\neq j$, can be written as \cite{walschaers2016efficient}
\begin{equation}
\begin{aligned}
C_{ij} &= \langle \hat {n}_i \hat {n}_j \rangle - \langle \hat {n}_i \rangle \langle \hat {n}_j \rangle = \\ 
&= \sum_{a \neq b} |\langle \chi_a | \chi_b \rangle|^2 U_{ia} U_{jb} \overline {U}_{ib} \overline {U}_{ja} - \sum_{k} |U_{ik}|^2|U_{jk}|^2.
\label{eq:two_mode_corr}
\end{aligned}
\end{equation}
The conservation of the total number of photons allows us to express the average variance in terms of these correlators. Defining $\hat{N}=\sum_i \hat{n}_i$ as the total photon number operator, we have from photon number conservation that 
\begin{align}
    \langle \hat{N}^2\rangle - \langle \hat{N}\rangle^2=0,     
\end{align}
which leads to
\begin{equation}
\sum_{i} \langle \hat {n}_i^2 \rangle - \langle \hat {n}_i \rangle^2 = - \sum_{i \neq j} \langle \hat {n}_i \hat {n}_j \rangle - \langle \hat {n}_i \rangle \langle \hat {n}_j \rangle,
\label{eq:sum_correlators}
\end{equation}
While a single two-mode correlator $C_{ij}$ was considered previously as a semi-device independent witness of indistinguishability \cite{vandermeer2021witness}, it can be seen from the equation above that the average variance takes into account all possible two-mode correlators in an unbiased way. This leads to better semidevice-independent guarantees on indistinguishability for unbiased as well as Haar random interferometers as demonstrated in the main text.

Using Eqs.~\eqref{eq:sum_correlators} and~\eqref{eq:two_mode_corr}, we can establish the following equality
\begin{align}
\sigma&=\frac{1}{n}\sum_{i} \langle \hat {n}_i^2 \rangle - \langle \hat {n}_i \rangle^2 \\&= -\frac{1}{n}\sum_{a \neq b} |\langle \chi_a | \chi_b \rangle|^2 \underbrace {\sum_{i \neq j} U_{ia} U_{jb} \overline {U}_{ib} \overline {U}_{ja}}_A + \nonumber\\
& ~~~~+ \frac{1}{n}\sum_k \underbrace {\sum_{i \neq j} |U_{ik}|^2 |U_{jk}|^2}_B.
\label{ap:sigma1}
\end{align}
\noindent The terms $A$ and $B$ can be simplified in the following way
\begin{equation}
\begin{aligned}
A &= - \sum_i U_{ia} U_{ib} \overline {U}_{ia} \overline {U}_{ib} + \sum_{i,j} U_{ia} \overline {U}_{ib} U_{jb} \overline {U}_{ja} = \\
&= - \sum_i |U_{ia}|^2|U_{ib}|^2 + \delta_{a,b}\delta_{a,b},
\end{aligned}
\end{equation}
and
\begin{equation}
B = - \sum_i |U_{ik}|^4 + \sum_{i} |U_{ik}|^2 \sum_j |U_{jk}|^2 = 1 - \sum_i |U_{ik}|^4. 
\end{equation}
\noindent Plugging in these simplifications in Eq.~\eqref{ap:sigma1}, we obtain the expression for $\sigma$ shown in the main text and in Eq.~\eqref{ap:exp_var}.

Furthermore, by considering a balanced interferometer such as the Fourier transform, where all matrix elements of the unitary are equal to $\frac {1}{\sqrt {n}}$ where $n$ is the number of modes, we obtain: 

\begin{equation} 
\sigma^{bal} = 1 + \sum_{a \neq b} \frac {|\langle \chi_a| \chi_b \rangle|^2}{n^2} - \frac {1}{n} =  1 - \frac {1}{n} + \frac {2}{n^2} \sum_{a>b} |\langle \chi_a | \chi_b \rangle|^2\,.
\end{equation}

In particular, for the three-mode balanced tritter implemented experimentally (see main text)  we have:

\begin {equation}
\begin{aligned}
\sigma^{bal}  &= \frac {2}{3} + \frac {2}{9}\left( |\langle \chi_1| \chi_2 \rangle|^2 + \right. \\
&+ \left. |\langle \chi_2| \chi_3 \rangle|^2 + |\langle \chi_1| \chi_3 \rangle|^2 \right). \label{eq:s3dqft}
\end{aligned}
\end {equation}

\subsection {Maximizing $\sigma$ for indistinguishable photons}\label {fdopt}

In this section we show that any balanced interferometer on $n$ modes maximizes $\sigma$ for perfectly indistinguishable photons; in particular, the Fourier interferometer maximizes $\sigma$. In the fully indistinguishable case all Gram matrix elements equal $1$, i.e. $\forall_{a,b}: \langle \chi_a | \chi_b \rangle = 1 $, which leads to
\begin{equation}
\begin{aligned}
\sigma &= 2 - \frac {2}{n} \sum_{a,i} |U_{ia}|^4. 
\end{aligned}
\end{equation}

\noindent Therefore, the maximization of $\sigma$ reduces to the minimization of $\sum_{a,i} |U_{ia}|^4$. We demonstrate that this term is lower-bounded by $1$ and that this lower bound is attained exactly for balanced interferometers, i.e. $|U_{ij}|^2 = \frac {1}{n} \forall i,j$, as is the case for the Fourier transform.

To demonstrate this we define the doubly-stochastic matrix $P_{ij}$ = $|U_{ij}|^2$ with $\sum_{i} P_{ij} =\sum_{j} P_{ij} = 1$. Using a known inequality between 1-norm and 2-norm, we have that:
\begin {equation}
\sqrt {\sum_i P_{ij}^2} \ge \frac {1}{\sqrt {n}} \sum_{i} P_{ij} =1 \implies \sum_j P_{ij}^2 \ge \frac {1}{n}. 
\end {equation}

\noindent Hence $\sum_{i,j} P_{ij}^2 \ge 1$. Moreover, it is easy to see that the bound is attained if $P_{ij}=1/n$. 

A similar reasoning can be taken for the fully distinguishability scenario, where all off-diagonal elements of the Gram matrix are $0$, i.e.  $\langle \chi_a | \chi_b \rangle = 0$,   $\forall {a \neq b}$. Here all the terms related to $a\neq b$ vanish leaving the expression just as
\begin{equation}
\sigma^{d} =1 - \frac {1}{n} \sum_{ik} |U_{ik}|^4, 
\end {equation}
which is also maximized for balanced interferometers. 

\section{Lower bounds for overlaps using the $n=3$ balanced interferometer}\label{ap:bound_Delta_min}

For the 3-mode Fourier transform, in Eq. (\ref{eq:s3dqft}) we found an expression for the average variance $\sigma$ of the photon number at the output modes in terms of two-photon overlaps $\Delta_i$ ($\Delta_1 = \Delta_{AB}, \Delta_2 = \Delta_{BC}, \Delta_3 = \Delta_{AC}$):

\begin{equation}
\sigma^{bal} = \frac {2}{3} + \frac {2}{9} \left(\sum_{i=1}^3 \Delta_i \right).
\end{equation}

\begin{figure}[ht!]
    \centering
    \includegraphics[width=0.49\textwidth]{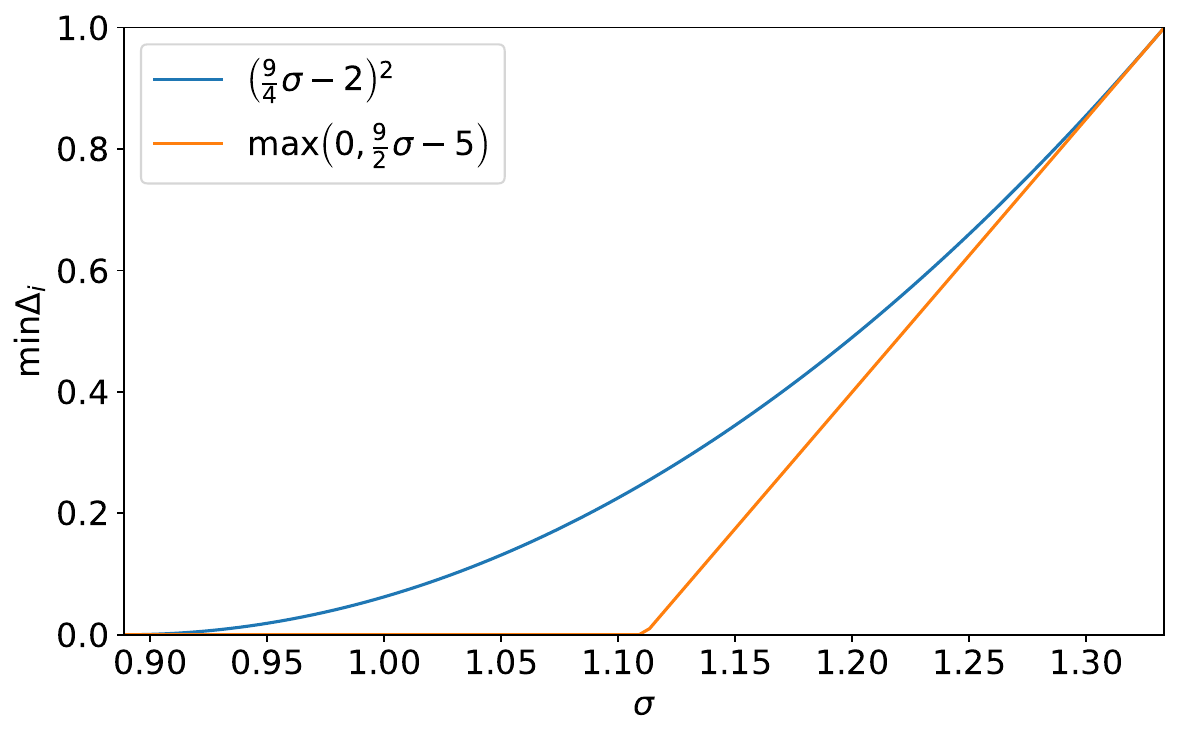}
    \caption{\textbf{Bounds on the two-photon overlaps.} Two lower bounds for the smallest overlap of 3 states, as a function of the observed average photon number variance $\sigma$. Blue curve is the best lower bound, obtained from the results of \cite{GalvaoB20}. The blue lower bound is non-trivial for $\sigma >8/9$, whereas the red lower bound is non-trivial only when $\sigma > 10/9$.}
    \label{fig:lbmin}
\end{figure}

In what follows we show how to obtain a lower bound for all overlaps using $\sigma$. This is a natural task when using $\sigma$ to certify the pairwise indistinguishability of a multiphoton source. We start by writing $\sum_i \Delta_i$ as a function of the measured $\sigma$:

\begin{equation}
\sum_i \Delta_i=\frac{9}{2}\sigma-3 . \label{eq:deltasigma}
\end{equation}
So if
\begin{equation}
\sigma >10/9 \rightarrow \sum_i \Delta_i >2 \rightarrow \min_i(\Delta_i)>0.
\end{equation}
More precisely, when $\sigma>10/9$, $\sigma$  gives a non-trivial lower bound for the smallest two-state overlap:
\begin{equation}
\min_i \Delta_i \ge \sum_i \Delta_i - 2=\frac{9}{2}\sigma-5. \label{eq:lb1}
\end{equation}
We can use the measured value of $\sigma$ in a Fourier interferometer to obtain better lower bounds for all overlaps $\Delta_i$. We recall that 3 quantum states have 3 overlaps $\Delta_i$ that satisfy the following inequality \cite{GalvaoB20}:
\begin{equation}
\sum_i \Delta_i -2\sqrt{\Delta_1 \Delta_2 \Delta_3}\le 1.
\end{equation}
Using Eq. (\ref{eq:deltasigma}) to write $\sum_i \Delta_i$ as a function of the variance $\sigma$, we get:
\begin{equation}
\sqrt{\Delta_1 \Delta_2 \Delta_3} \ge \frac{9}{4}\sigma -2 \Rightarrow \Delta_1 \Delta_2 \Delta_3 \ge \left(\frac{9}{4}\sigma-2 \right)^2.
\end{equation}
The product of all $\Delta_i$ is a lower bound for each individual $\Delta_i$, in particular for $\min_i \Delta_i$, so:
\begin{equation}
\min_i \Delta_i \ge \Delta_1 \Delta_2 \Delta_3 \ge \left(\frac{9}{4}\sigma-2 \right)^2.
\end{equation}
The inequality above give us a better (larger) lower bound for all overlaps, in terms of the observed variance $\sigma$, than Eq. (\ref{eq:lb1}) does. In particular, it guarantees a non-trivial lower bound when $\sigma >8/9$, whereas Eq. (\ref{eq:lb1}) only does that for $\sigma >10/9$. In Fig. \ref{fig:lbmin} we plot the two lower bounds as a function of $\sigma$.

\section{Semi-device independent bounds on the average indistinguishability}
\label{ap:bounds_average_delta}
To simplify the expression for the average photon number variance in Eq.~\eqref{eq:sigma_U} it is useful to introduce some notation. We define $\Delta_{ab}=|\langle \chi_a | \chi_b \rangle|^2$ as the $n\times n$ matrix containing all the overlaps. Moreover we define the doubly stochastic matrix $P_{ab}=|U_{ab}|^2$ associated to the unitary $U$, as well as the (also doubly stochastic) matrix $Q=P^{T} P$. The expression for $\sigma$ can now be written as 
\begin{align}
    \sigma&= 1+ \frac{1}{n} \sum_{a\neq b} \Delta_{ab}Q_{ab}-\frac{1}{n}\text{tr}(Q). \\
\end{align}
For fully distinguishable photons we obtain 
\begin{align}\label{eq:sigma_d}
    \sigma^d&= 1-\frac{1}{n}\text{tr}(Q).
\end{align}
The difference between the measured values of $\sigma$, for some distinguishability scenario, and $\sigma^d$ gives us
\begin{equation}
    \sigma-\sigma^d= \frac{1}{n} \sum_{a\neq b} \Delta_{ab}Q_{ab}. 
\end{equation}

Unless $U$ is the identity matrix, the difference between $\sigma$ and $\sigma^d$ will be positive and will give some information about the overlap matrix $\Delta$. To quantify this difference, let us rewrite the equation as follows
\begin{align}
    \sigma-\sigma^d&= \frac{1}{n} \left(\text{tr}(\Delta Q)-\text{tr}(Q)\right).
\end{align}
Using Eq.~\eqref{eq:sigma_d}, we have 
\begin{align}\label{eq:sigma_difference}
    \sigma -2\sigma^d +1 = \frac{\text{tr}(\Delta Q)}{n}.
\end{align}
Since $Q$ and $\Delta$ are positive-semidefinite matrices, we can use the inequality $\text{tr}(\Delta Q)\leq ||\Delta|| \text{tr}{(Q)}$, where $||\Delta||$ denotes the spectral norm (largest eigenvalue) of $\Delta$. This leads to the following inequality 
\begin{equation}\label{eq:ineq_normDelta}
    \frac{||\Delta||}{n}\geq \frac{\sigma - 2 \sigma^d +1}{n (1- \sigma^d )}. 
\end{equation}
While $||\Delta||/n$ itself could be seen as a measure of indistinguishability, our aim here is to use this bound to obtain guarantees on the average indistinguishability $\bar{\Delta}$ from Eq.~\eqref{eq:av_Delta}. To do so, we first use the fact that $\Delta=S\odot S^{T}$, where $S$ is the Gram matrix $S_{ab}=\langle \chi_a | \chi_b \rangle$, where $\odot$ denotes the Hadamard (entrywise) product. It follows that $||\Delta||\leq ||S||$ \cite{horn1995norm} which, together with Eq.~\eqref{eq:ineq_normDelta}, implies that
\begin{equation}\label{eq:ineq_normS}
    \frac{||S||}{n}\geq \frac{\sigma - 2 \sigma^d +1}{n (1- \sigma^d )}. 
\end{equation}
In order to write an inequality in terms of the average indistinguishability, we can use the fact that $||S||_F= \sqrt{\sum_{a,b} \Delta_{ab}}\geq ||S||$, where $||S||_F$ is the Frobenius norm, which can be written in terms of $\bar{\Delta}$ as
\begin{align}
\frac{||S||_F}{n}= \sqrt{\frac{n(n-1)}{n^2}\bar{\Delta}+1/n}.
\end{align}
This expression, together with the bound $||S||_F\geq ||S||$ and Eq.~\eqref{eq:ineq_normS}, implies that 
\begin{equation}
    \bar{\Delta}\geq \frac{1}{n(n-1)}\left(\frac{\sigma-2\sigma^d+1}{1-\sigma^d}\right)^2-\frac{1}{n-1}.
\end{equation}

\section{Integrated photonic processor}
\label{ap:processor}
The experiment has been performed by using an 8-mode fully-programmable IPP. The fabrication of the IPP is based on FLW as described in \cite{corrielli2018symmetric}. The single-mode waveguides were inscribed in an alumino-borosilicate glass substrate (Corning Eagle XG); the device features a mesh of 28 MZIs in a rectangular configuration, allowing the implementation of any unitary matrix transformation in 8 modes. The waveguides were optimized for a wavelength of \SI{928}{\nano \meter} and the designs of the interferometers and the mesh as a whole have been optimized for higher integration density as detailed in \cite{albiero2022toward}. On top of each MZI, a pair of resistive heating elements has been fabricated through a new two-metal photolithographic approach \cite{albiero2022toward}. These elements allow the control of each MZI's operation through phase shifting via the thermo-optic effect, with a dissipated power needed to achieve a $2\pi$ shift of about \SI{35}{\milli \watt} on average per MZI. Finally, fiber arrays are glued with UV-curing resin to both the input and output ports of the device, allowing easy interfacing with the photon sources and detectors. The integrated interferometer has a planar footprint of about $1 \times \SI{80}{\square \milli \meter}$ and total insertion losses below \SI{3}{\decibel}.

\begin{figure*}
    \centering
    \includegraphics[width=0.99\textwidth]{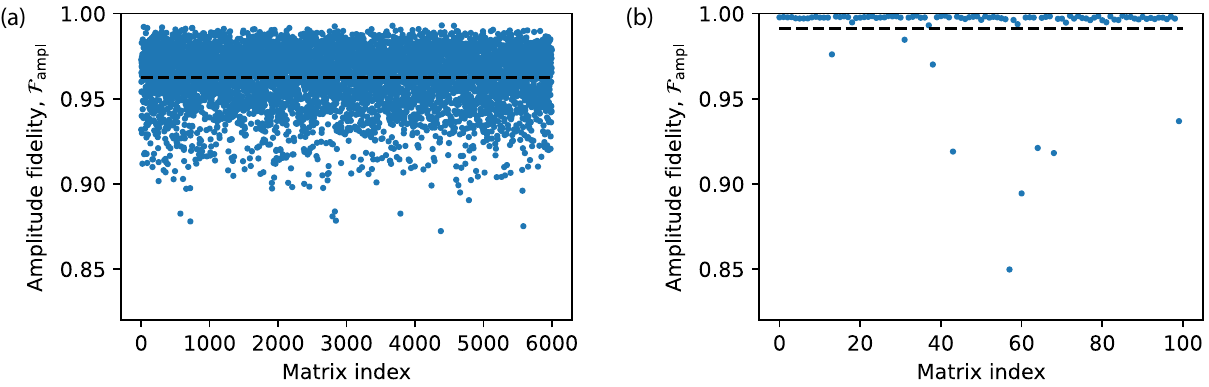}
    \caption{\textbf{Implementation of random Haar unitary matrices on the 8 mode IPP.} The fidelities over (a) 6000 random Haar matrices before the optimization algorithm and (b) 100 random Haar matrices after the optimization algorithm. The average fidelity is marked with a horizontal dashed line.}
    \label{fig:milan_fidelities}
\end{figure*}

The model we employ for the relationship between the vector of currents $I$ and the vector of MZI phases $\phi$ is as follows:
\begin{equation}
	\phi = \phi_0 + A_1 I^2 + A_2 I^4,
\end{equation}
where $\phi_0$ is the vector of initial phases, while $A_1$ and $A_2$ are matrices respectively representing all the terms that are linear in the dissipated power at each resistive element including the thermal cross-talk terms and representing the terms that depend on the square of the dissipated power. A calibration procedure was employed to retrieve the $\phi_0$, $A_1$, and $A_2$ arrays, in order to implement the required unitary transformations for this work. The calibration consists of a node isolation algorithm where coherent light is injected at the input of each individual MZI \cite{pentangelo2024high}. The performance of the calibration has been measured with the \emph{amplitude fidelity} figure of merit:
\begin{equation}
    \mathcal{F}_\text{ampl} = \frac{1}{8} \sum_{ij} |(U_t)_{ij}| |(U_m)_{ij}|,
\end{equation}
where $U_t$ and $U_m$ are the target unitary matrix and the implemented unitary matrix, respectively. Through this calibration, it was possible to achieve an average fidelity of 0.962 over a sample of 6000 random Haar matrices, as reported in Fig. \ref{fig:milan_fidelities} (a).

After this preliminary calibration step, the device's operation was further optimized algorithmically. In particular, the predicted currents from a large sample size of 6000 random Haar matrices were calculated taking the initial calibration as a starting point. A machine learning algorithm was employed using the data as a training set, to further tune the calibration parameters in order to maximize the amplitude fidelity. This algorithm optimized the model parameters, addressing imbalances in the couplers in each MZI, non-idealities in the manual calibration process, and fabrication tolerances. After this step, the average fidelity was increased to 0.991 as reported in Fig. \ref{fig:milan_fidelities} (b).

\section{Characterization of tritter unitary transformation} 
\label{ap:interferometer}

In this Section we describe the procedure employed to reconstruct the unitary matrix $U$ describing the balanced tritter transformation, which was experimentally realized using our integrated photonic device. In the linear optics formalism, one can characterize the action of an $n$-mode integrated interferometer as a $n\cross n$ unitary matrix mapping the creation mode operators $\vec{a}^\dag$ into output mode creation operators $\vec{b^\dag}$:
\begin{equation}
    \mqty(b^\dag_0 \\ b^\dag_1 \\ \dots \\ b^\dag_n) = U \mqty(a^\dag_0 \\ a^\dag_1 \\ \dots \\ a^\dag_n).
\end{equation}
As described in Ref. \cite{laing2012super}, in order to reconstruct the elements $U_{\alpha \beta} = |U_{\alpha \beta}|e^{i \phi_{\alpha \beta}}$ of $U$ it is in principle sufficient to estimate the following quantities:
\begin{equation}
    \mathcal{V}_{ij,mn} = 1 - \frac{P_{ij,mn}^I}{P_{ij,mn}^D},
    \label{eqn:Vij}
\end{equation}
where $P_{ij,mn}^I$ gives the probability of recovering a coincidence click on outputs $(i,j)$ when injecting identical photons in inputs $(m,n)$, while $P_{ij,mn}^D$ identifies the analogous quantity computed with distinguishable photons, i.e. in absence of quantum interference. The quantity $\mathcal{V}_{ij,mn}$ depends on both the moduli and the phases of each involved unitary element, i.e. $(U_{im},U_{in},U_{jm},U_{jn})$. An over-complete characterization of the unitary $U$ requires to inject a pair of photons in each possible combination of input modes. Formally, the elements present in Eq. \eqref{eqn:Vij}, are associated to the $2\times 2$ sub-block of the unitary matrix $U$:
\begin{equation}
    \bf{U}_{ijmn} = \mqty(U_{im} & U_{in} \\ 
          U_{jm} & U_{jn}).
\end{equation}
Indeed, the output coincidence probability for indistinguishable photons is related to the square moduli of the permanent of $\bf{U}_{ijmn}$:
\begin{equation}
\begin{aligned}
    P_{ij,mn}^I &= |\text{Per}({\bf{U}_{ijmn}})|^2 = |U_{im}U_{jn} + U_{in}U_{jm}|^2 = \\
    &= |U_{im}U_{jn}|^2 + |U_{in}U_{jm}|^2 + \\ 
    &+ 2|U_{im}U_{jn}U_{in}U_{jm}| \cos(\phi_{im} + \phi_{jn} - \phi_{in} - \phi_{jm}),
\end{aligned}
\end{equation}
while for completely distinguishable photons the output coincidence probability is given by the permanent of the moduli square matrix $|{\bf{U}_{ijmn}}|^2$:
\begin{equation}
    P_{ij,mn}^D = \text{Per}(|{\bf{U}_{ijmn}}|^2) = |U_{im}U_{jn}|^2 + |U_{in}U_{jm}|^2.
\end{equation}

In our optical setup, based on a QD single photon source paired with a time-to-spatial demultiplexing setup, a direct estimation of the quantities described in Eq. \eqref{eqn:Vij} requires taking into account the presence of the time-to-spatial demultiplexed setup, and the resulting non-unitary indistinguishability amongst photon pairs. Then, similarly to Eq. \eqref{eqn:Vij}, one can consider the following quantity:
\begin{equation}
    R_{ij,mn} = \frac{A^{\tau = 0}_{ij,mn}}{A^{\tau = \pm T}_{ij,mn}},
    \label{eqn:ratios}
\end{equation}
where the terms in the numerator and the denominator are respectively related to the areas of the zero($T$)-delay  coincidence peaks in a typical correlation histogram measured at outputs $(i,j)$ when injecting a stream of photon pairs in inputs $(m,n)$. Here, $T$ represent the 12 ns separation between subsequent peaks due to the 79 MHz excitation of the QD. Indeed, the two-photon input state in our optical interferometer can be described by the overlap $\omega$, where this parameter takes into account the partial distinguishability amongst the input photons' wavepackets. Then, when considering the zero-delay coincidence peak $A^{\tau = 0}_{ij,mn}$, it can be shown that this quantity is proportional to:
\begin{equation}
\begin{aligned}
    A^{\tau = 0}_{ij,mn} & \propto P^{\tau = 0}_{ij,mn} = \omega P_{ij,mn}^I  + (1-\omega) P_{ij,mn}^D = \\ & = |U_{im}U_{jn}|^2 + |U_{in}U_{jm}|^2 + \\
    &+ 2 \omega |U_{im}U_{jn}U_{in}U_{jm}| \cos(\phi_{im} + \phi_{jn} - \phi_{in} - \phi_{jm}).
\end{aligned}
\end{equation}
Conversely, a contribution to the $T$-delay coincidence peak will be given by two photons separated by $T$ entering the interferometer in inputs $\{(m_0,m_T), (n_0,n_T), (m_0,n_T), (n_0,m_T)\}$, thus leading to:
\begin{equation}
     A^{\tau = T}_{ij,mn} \propto |U_{im}U_{jm}|^2 + |U_{in}U_{jn}|^2 + |U_{im}U_{jn}|^2 + |U_{in}U_{jm}|^2.
\end{equation}
Note that in such a way, the ratios $R_{ij,mn}(U,\omega)$ will be a function of both the two photon overlap $\omega$ and the implemented unitary transformation $U$. Thus, by measuring the ratios in Eq. \eqref{eqn:ratios} for each possible combination of inputs $(m,n)$ and outputs $(i,j)$, together with the HOM visibilities, we can solve the following optimization problem to recover the experimentally implemented $\tilde{U}_{3}$ transformation:
\begin{equation}
    \min_{U} \sum_{ijmn} \frac{[\tilde{R}_{ijmn} - R_{ijmn}(U,\tilde{V})]^2}{\sigma^2(\tilde{R}_{ijmn})},
    \label{eqn:optimi}
\end{equation}
where $\tilde{R}_{ijmn}$ are the experimentally measured ratios, $\sigma^2(\tilde{R}_{ijmn})$ are the associated experimental error and $R_{ijmn}(U,\tilde{V})$ are the theoretical predictions for $\omega = \tilde{V}$, being $\tilde{V}$ the experimental two-photon visibility. Note that, generally, a 3x3 complex unitary matrix will depend only on 8 free parameters. Thus we can exploit the parametrization presented in Ref. \cite{PhysRevD.38.1994} to decrease the complexity of the optimization problem. Overall, we obtain the reconstructed matrix element moduli:
\begin{equation}
\vert \tilde{U}_3 \vert = 
\begin{pmatrix}
0.611 \pm 0.001 & 0.552 \pm 0.001 & 0.567 \pm 0.001 \\
0.559 \pm 0.001 & 0.598 \pm 0.001 & 0.575 \pm 0.001 \\
0.561 \pm 0.001 & 0.581 \pm 0.001 & 0.590 \pm 0.001
\end{pmatrix},
\end{equation}
with phases
\begin{equation}
\arg(\tilde{U}_3) = 
\begin{pmatrix}
0 & 0 & 0 \\
0 & 2.128 \pm 0.004 & -2.107 \pm 0.004\\
0 & -2.087 \pm 0.004 & 2.155 \pm 0.003
\end{pmatrix},
\end{equation}
where, without loss of generality, the complex phases in both the first column and row of the matrix can be constrained to be zero.
The reconstructed matrix $\tilde{U}_3$ reached a fidelity of $F = \vert \mathrm{Tr}(U_3 \tilde{U}_3^{\dag}) \vert/3 > 0.999$ w.r.t. the ideal tritter unitary $U_3$. Here, standard errors have been computed as a mean over $\approx 500$ successfully converged instances of the problem in Eq. \eqref{eqn:optimi}, where the input $R^{\exp}_{ijmn}$ are statistically sampled from a normal distribution.

\section{Testing the Gram-matrix phase}
\label{ap:cyclic}

In the main text, we considered the measurement of the different tests by considering positive real-valued Gram-matrices of the form:
\begin{equation}
\label{eq:Sk_si}
S_{k} = 
\begin{pmatrix}
1 & \sqrt{\tilde{\Delta}^{k}_{ab}} & \sqrt{\tilde{\Delta}^{k}_{ac}} \\
\sqrt{\tilde{\Delta}^{k}_{ab}} & 1 & \sqrt{\tilde{\Delta}^{k}_{bc}}\\
\sqrt{\tilde{\Delta}^{k}_{ac}} & \sqrt{\tilde{\Delta}^{k}_{bc}} & 1
\end{pmatrix}.
\end{equation}
To justify this assumption on the tested input state configuration, we verified experimentally that the output states generated from our source are well represented by this form of the Gram-matrix. This can be done by exploiting the class of cyclic interferometers employed in Ref. \cite{Pont22} as a mean to measure genuine multiphoton indistinguishability. Let us then consider a 6-mode cyclic interferometer described by the matrix:
\begin{equation}
U_{cyc}(\alpha) = \frac{1}{2} \begin{pmatrix}
1 & -1 & 1 & 1 & 0 & 0\\
1 & -1 & -1 & -1 & 0 & 0\\
e^{\imath \alpha} & e^{\imath \alpha} & 0 & 0 & 1 & -1\\
e^{\imath \alpha} & e^{\imath \alpha} & 0 & 0 & -1 & 1\\
0 & 0 & 1 & -1 & 1 & 1\\
0 & 0 & 1 & -1 & -1 & -1
\end{pmatrix},
\end{equation}
where $\alpha$ is a tunable phase shift. This matrix belongs to the same class of cyclic interferometers of \cite{Pont22}, up to mode swaps and to a different complex phase choice for the elementary beam-splitter unitary cell. 

Let us then consider the scenario with three input photons, described by three internal states $\vert \chi_{a} \rangle$, $\vert \chi_{b} \rangle$, and $\vert \chi_{c} \rangle$. In this case, the general form of the Gram-matrix can be written as:
\begin{equation}
S = \begin{pmatrix}
1 & \sqrt{\Delta_{ab}} & \sqrt{\Delta_{ac}} \\
\sqrt{\Delta_{ab}} & 1 & \sqrt{\Delta_{bc}} e^{\imath \varphi} \\
\sqrt{\Delta_{ac}} & \sqrt{\Delta_{bc}} e^{- \imath \varphi} & 1
\end{pmatrix}.
\label{eq:GM_general}
\end{equation}
Let us consider the case where the three photons are injected in input ports $[1,3,5]$ of $U_{cyc}$. The input state can be thus described as $a^{\dag}_{1} b^{\dag}_{3} c^{\dag}_{5} \vert 0 \rangle$, where the operators $\{ a^{\dag}_{i}, b^{\dag}_{j}, c^{\dag}_{k} \}$ label respectively the creation operators for a photon in internal states $\vert \chi_{a} \rangle$, $\vert \chi_{b} \rangle$, $ \vert \chi_{c} \rangle$ and input modes $i,j,k$ respectively. We consider the subset of output modes corresponding to the relevant configurations identified by \cite{Pont22}, namely the sets $s_+ = \{[1,3,5]$, $[1,4,5]$, $[2,3,6]$, $[2,4,5]\}$ and $s_{-} = \{[1,3,6]$, $[1,4,5]$, $[2,3,5]$, $[2,4,6]\}$. We then transform the creation operators of the input state according to the action of the interferometer. This is performed by considering that according to the adopted notation the input-output relation is $\hat{a}^{\dag}_{i} \rightarrow \sum_{j} [U^{t}_{cyc}(\alpha)]_{ij} \hat{a}^{\dag}_{j}$, where $t$ stands for the matrix transpose, and analogously for $\hat{b}^{\dag}_{i}$ and $\hat{c}^{\dag}_{i}$. The input state $a^{\dag}_{1} b^{\dag}_{3} c^{\dag}_{5} \vert 0 \rangle$ then evolves into:
\begin{equation}
\begin{aligned}
\vert \psi_{cyc}(\alpha) \rangle &= \frac{1}{8} (\hat{a}^{\dag}_{1} + \hat{a}^{\dag}_{2} + e^{\imath \alpha} \hat{a}^{\dag}_{3} + e^{\imath \alpha} \hat{a}^{\dag}_{4}) \times \\ &\times(\hat{b}^{\dag}_{1} - \hat{b}^{\dag}_{2} + \hat{b}^{\dag}_{5} + \hat{b}^{\dag}_{6}) \times (\hat{c}^{\dag}_{3} - \hat{c}^{\dag}_{4} + \hat{c}^{\dag}_{5} - \hat{c}^{\dag}_{6}) \vert 0 \rangle.
\end{aligned}
\end{equation}
Let us first focus on output configuration $[1,3,5] \in s_+$. We first select the (unnormalized) portion of the output state containing one operator for each output mode: 
\begin{equation}
\vert \psi_{1,3,5} \rangle = \frac{1}{8} \left(\hat{a}^{\dag}_{1} \hat{b}^{\dag}_{5} \hat{c}^{\dag}_{3} + e^{\imath \alpha} \hat{a}^{\dag}_{3} \hat{b}^{\dag}_{1} \hat{c}^{\dag}_{5} \right) \vert 0 \rangle.
\end{equation}
The probability of this output configuration can be then obtained by evaluating the overlap $\langle \psi_{1,3,5} \vert \psi_{1,3,5} \rangle$. By repeating the same calculation for all states belonging to sets $s_+$ and $s_-$, the output probabilities are found to be:
\begin{equation}
P_{\pm}(\alpha, \varphi) = \frac{1}{32} \left[1 \pm \sqrt{\Delta_{ab}} \sqrt{\Delta_{bc}} \sqrt{\Delta_{ac}} \cos(\alpha + \varphi)\right],
\end{equation}
where the $\pm$ sign depends on whether the output configuration belongs to set $s_+$ or $s_-$.

\begin{figure}[ht!]
\centering
\includegraphics[width=0.49\textwidth]{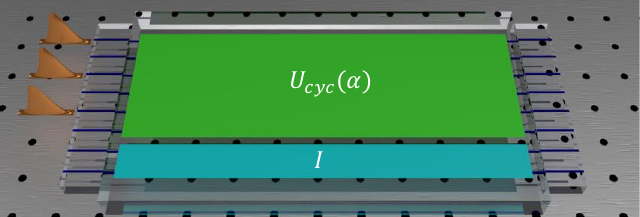}
\caption{\textbf{Device programming for the cyclic interferometers.} The processor was also programmed to act as a cyclic interferometer $U_{cyc}(\alpha)$ over the first six modes to test the assumption of a real-valued Gram matrix. In both schemes, $I$ corresponds to the identity over the modes.}
\label{fig:cyclic_circuit}
\end{figure}

\begin{figure}[ht!]
\centering
\includegraphics[width=0.49\textwidth]{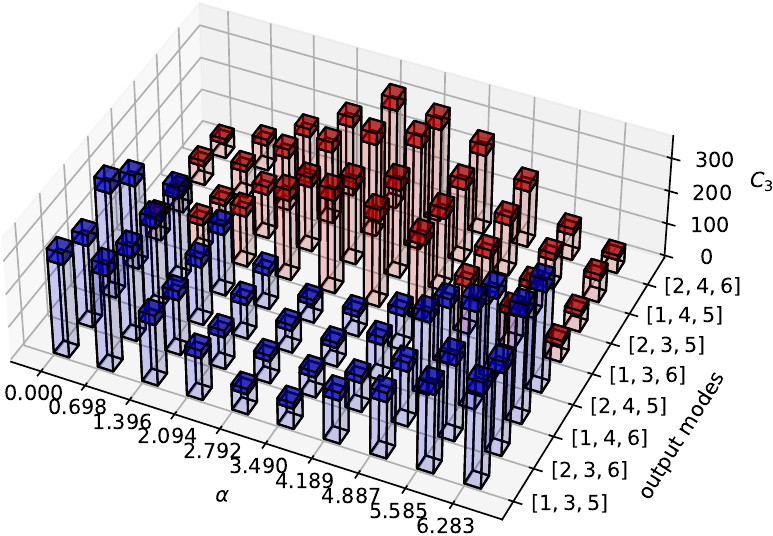}
\caption{\textbf{Measurements with a 6-mode cyclic interferometer.} Measured three-fold coincidences $C_{3}$ obtained by programming the IPP to act as $U_{cyc}(\alpha)$ for input modes $[1,3,5]$ and output configurations in $s_+$ (blue bars) and $s_-$ (red bars), for different values of $\alpha$. The darker regions in the bars correspond to the 1-$\sigma$ measured interval for the output three-fold coincidences according to the Poissonian statistics of the detected events.}
\label{fig:c1_barplot}
\end{figure}

\begin{figure}[ht!]
\centering
\includegraphics[width=0.49\textwidth]{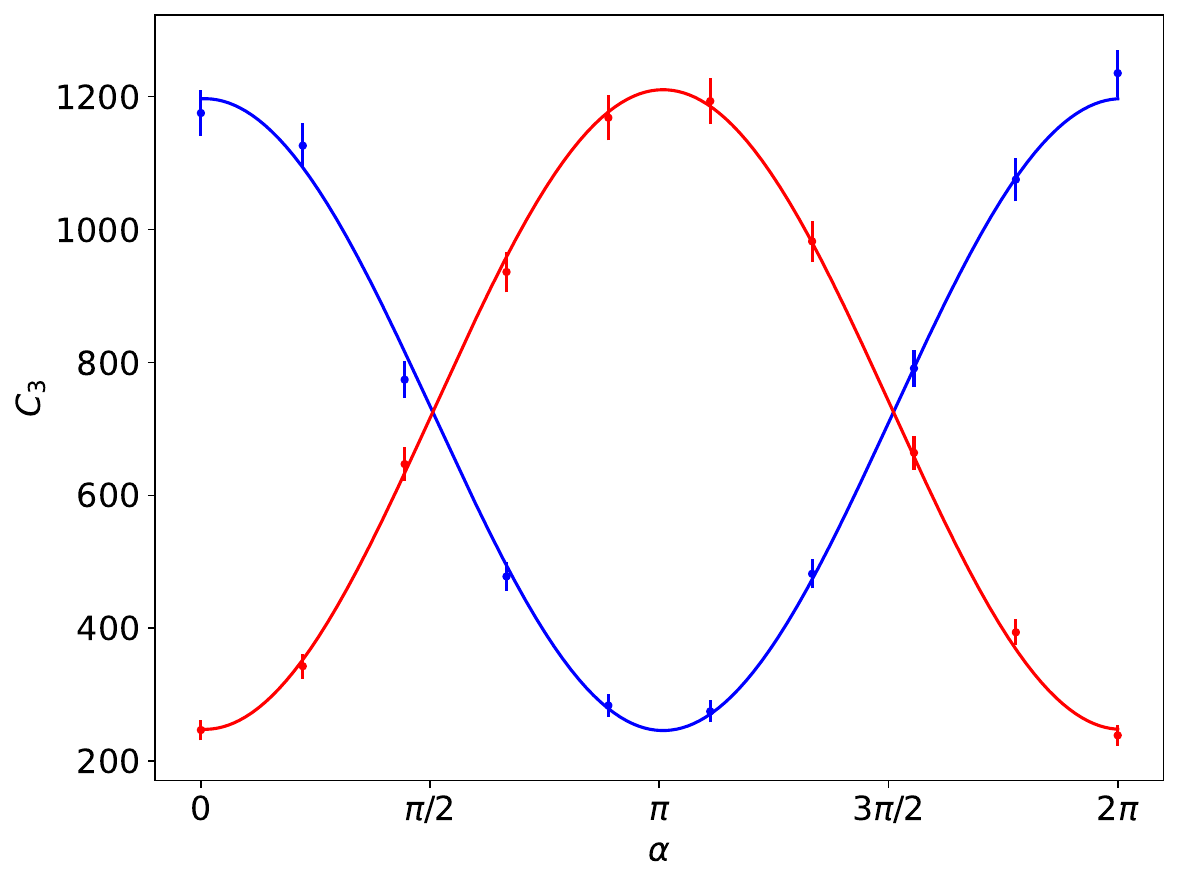}
\caption{\textbf{Estimation of the three-photon Gram-matrix phase $\varphi$.} Three-fold coincidences $C_3$ at the output of the interferometer programmed as $U_{cyc}(\alpha)$ from three photon in input modes $[1,3,5]$, grouped in output sets $s_+$ (blue) and $s_-$ (red). Solid lines are best-fit with a sinusoidal model, with the constraint that the two curves share the same phase-offset. Error bars are due to the Poissonian statistic of the detected events.}
\label{fig:c1_fit}
\end{figure}

The value of the complex phase $\varphi$ of the Gram-matrix can be thus retrieved by injecting the three photons on input ports $[1,3,5]$ measuring the output probabilities, and by measuring the probabilities of output configurations in sets $s_+$ and $s_-$ after programming the interferometer to act as $U_{cyc}(\alpha)$ for different values of $\alpha$ (see Fig. \ref{fig:cyclic_circuit}). The results of this measurement for scenario $S_{A}$ are shown in Fig. \ref{fig:c1_barplot} for 10 different values of $\alpha$ covering the $[0, 2\pi)$ interval. The value of $\varphi$ can be then obtained by grouping the measured output three-fold coincidences according to the $s_+$ and $s_-$ sets, and by fitting the grouped data with a sinusoidal function as a function of $\alpha$, given that $\varphi$ corresponds to the phase offset of the patterns. In our case (see Fig. \ref{fig:c1_fit}), we found $\varphi = -0.03(2)$, which is compatible with $\varphi_0 = 0$ within 2 standard deviations. This approach, and the corresponding experimental verification, justifies the adoption of positive real-valued Gram-matrices in the main text.

An independent verification on the assumption of positive real-valued Gram-matrices can be also obtained by direct analysis of the measured 3-photon distribution after the tritter transformation, assuming knowledge on the two-photon overlaps. Let us then consider again scenario $S_{A}$, corresponding to measured values of the overlaps $\tilde{\Delta}^{A}_{ab} = 0.875(4)$, $\tilde{\Delta}^{A}_{ac} = 0.874(2)$ and $\tilde{\Delta}^{A}_{bc} = 0.848(2)$. We then consider the general Gauge-invariant form of the Gram matrix for 3 photons [Eq. (\ref{eq:GM_general})], leaving $\varphi$ as a free parameter. Let $\tilde{p}_{(n_1,n_2,n_3)}$ the experimentally measured probabilities after the tritter transformation $\tilde{U}_{3}$, where $(n_1, n_2, n_3)$ is one of the possible photon number occupation configurations on the output modes. Conversely, let $p_{(n_1,n_2,n_3)}(\varphi)$ be the expected value of the output probabilities for a Gram matrix having two-photon overlap equal to the measured values reported above, and value $\varphi$ for the phase. One can then define the following quantity:
\begin{equation}
e(\varphi) = \sum_{n_1,n_2,n_3} \frac{[\tilde{p}_{(n_1,n_2,n_3)} - p_{(n_1,n_2,n_3)}(\varphi)]^2}{\sigma^2(\tilde{p}_{(n_1,n_2,n_3)})},
\end{equation}
where $\sigma(\tilde{p}_{n_1,n_2,n_3})$ is the experimental uncertainty associated to $\tilde{p}_{n_1,n_2,n_3}$, and the sum extended over all possible 3-photon configurations $(n_1,n_2,n_3)$ where $n_i$ is the occupation number of mode $i$ thus satisfying $\sum_i n_i = 3$. 
This quantity represents the squared error between the expectations for a given value of $\varphi$, and the experimentally measured values, weighted by the experimental uncertainty. One can then retrieve the value of $\varphi$ by numerically minimizing this quantity. We have performed this optimization by considering as a model the closed forms of $p_{(n_1,n_2,n_3)}(\varphi)$ calculated in \cite{menssen2017distinguishability}, which corresponding to an ideal tritter with no multiphoton contributions from the source, and our detailed model including the main sources of noises. The value of $\varphi$ obtained with the two choices of the expectations $p_{(n_1,n_2,n_3)}(\varphi)$ after minimization of $e(\varphi)$ are found to be respectively $\varphi = 0.0053(1)$ and $\varphi = 0.0078(1)$, which are compatible with the measurement performed via the cyclic interferometer.

\section{Model of the experiment} 
\label{ap:model}

In this Section we provide a brief overview of the theoretical model for the experimental data, which follows a similar approach to the one reported in \cite{Pont22}. In the present implementation, we need to take into account different features of the platform: (i) the actual value of the photon indistinguishability, quantified by the Gram matrix $S$, (ii) multiphoton contributions from the source, (iii) the effective unitary transformation $\tilde{U}$ implemented by the integrated device and (iv) losses in the apparatus.

Regarding photon indistinguishability (i), this is taken into account by considering that the output probabilities are evaluated according to the theory reported in \cite{tichy2015sampling}, which explictly includes the overlap between photon states according to the Gram matrix $S$.

On a second note, we can include the effect of multiphoton emission (ii) by considering that on each time-bin (before the demultiplexing module), and thus on each spatial mode, there is a non-negligible probability that a second photon is emitted. This corresponds to write the input state on each mode as:
\begin{equation}
\rho = p_{0} \vert 0 \rangle \langle 0 \vert + p_{1} \vert 1 \rangle \langle \tilde{1} \vert + p_{2} \vert 1,\tilde{1} \rangle \langle 1,\tilde{1} \vert,
\end{equation}
where the $\tilde{1}$ corresponds to the presence of an additional noise photon, which for the employed quantum-dot source has zero-overlap with the principal photon \cite{Olli21}. The parameters $[p_{0}, p_{1}, p_{2}]$ can be retrieved, from the source brightness and from the measured $g^{(2)}(0)$, as $B \sim p_1 + p_2$ and $g^{(2)}(0) = 2 p_2/(p_1 + 2 p_2)^2$.

While the effective unitary $\tilde{U}$ (iii) can be directly included in the probability distribution calculation, the remaining feature to be taken into account is the presence of losses (iv). In our apparatus, we observe that losses are approximately balanced for each photon. We can model this effect by considering that a round of balanced losses commutes with passive linear optical circuits \cite{Oszm18}. Thus, we can approximate that all losses in our apparatus can be grouped and placed at the output of the source, where the transmission coefficient per photon $\eta_0$ is obtained as the product of all transmissivities of each element. Finally, we observe that in our model, we do not have to take into account unbalancements in the detection system efficiencies, since the measurement apparatus has been characterized and the output probabilities have been normalized accordingly.

\section{Data analysis}
\label{ap:data_analysis}

In this Section we provide more details on the analysis performed on the experimental data, reported in the main text.

\subsection{Pseudo Number Resolved detection}

The reconstruction of the bunching probabilities reported in the main text requires the capabilities of distinguishing the photon number at the output of each mode of the programmed 3-mode unitary transformation. To this end, we employed a probabilistic pseudo-photon number detection approach. More specifically, if one wants to detect up to 3 photons in a given output mode $i$ ($i =1,2,3$) with threshold detectors, such a mode can be divided among three auxiliary modes $i\alpha$ ($\alpha =1,2,3$) by means of an appropriate cascade of beam splitters. In our experiment, we employed the remaining layers of the IIP together with an external in-fiber single mode 50:50 beam splitter to perform such a balanced mode splitting. After this operation, the probability of detecting a photon exiting mode $i$ in the detection mode $i\alpha$ will be $P_{i\rightarrow i\alpha} = 1/3$. One can then compute the probability of observing an output configuration $\vec{n} = (n_1,n_2,n_3)$ with $\sum_{i} n_i = 3$. It is easy to show that, given a detection efficiency $\eta$ and by denoting as $\{n_1,n_2,n_3\}$ a valid permutation of three photons exiting 3 modes, the detection probabilities of an output configuration $\{n_1,n_2,n_3\}$ with the probabilities scheme reads:
\begin{eqnarray}
    P^{\text{det}}_{\{3,0,0\}} &=& 2\eta^3/9, \\
    P^{\text{det}}_{\{2,1,0\}} &=& 2\eta^3/3, \\
    P^{\text{det}}_{\{1,1,1\}} &=& \eta^3.
\end{eqnarray}
This detection probabilities act as an effective additional loss factor dependent on the output configuration. This means that, to estimate the full three-photon probability distribution, it is necessary have to normalize the measured three-fold counts $\tilde{N}_{(n_1, n_2, n_3)}$ as:
\begin{equation}
    N_{(n_1,n_2,n_3)} = \tilde{N}_{(n_1,n_2,n_3)}/P^{\text{det}}_{\{n_1,n_2,n_3\}},
    \label{eqn:correction}
\end{equation}
where $\tilde{N}_{(n_1,n_2,n_3)}$ if the effective measured number of events associated to configuration $(n_1,n_2,n_3)$. In such a way, with a total of 9 detectors, one can reconstruct the full probability distribution at the output of a generic transformation $U$:
\begin{equation}
    \tilde{p}_{(n_1,n_2,n_3)} =  \frac{N_{(n_1,n_2,n_3)}}{\sum_{n_1,n_2,n_3} N_{(n_1,n_2,n_3)}},
    \label{eqn:fullprob}
\end{equation}
where $n_1,n_2,n_3 \in \{0,1,2,3\}$ with $\sum_i n_i = 3$. Without loss of generality, the detection efficiency can be included in a direct estimation of the experimental mode-detection probabilities $\tilde{P}_{i\rightarrow i\alpha}$. This coefficients have been measured by setting $U = I$ and injecting the IIP with single photon states at each input mode $i$, measuring:
\begin{equation}
    \tilde{P}_{i\rightarrow i\alpha} = \frac{N_{i\alpha}}{\sum_j N_{i\alpha}},
\end{equation}
where $i,\alpha \in (1,2,3)$. This expression can be then used to measure the detection coefficients $P^{\text{det}}_{\{n_1,n_2,n_3\}}$ which we report in Table \ref{tab:efficiencies}.

\begin{table}[ht!]
    \centering
    \begin{tabular}{c|c}
    $(n_1,n_2,n_3)$ & $P^{\text{det}}_{\{n_1,n_2,n_3\}}$ \\
    \hline\hline
    (1, 1, 1) & 1.0 \\
    (2, 1, 0) & 0.666 \\
    (2, 0, 1) & 0.666 \\
    (1, 2, 0) & 0.6646 \\
    (0, 2, 1) & 0.6646 \\
    (0, 1, 2) & 0.6662 \\
    (1, 0, 2) & 0.6662 \\
    (3, 0, 0) & 0.2216 \\
    (0, 3, 0) & 0.2201 \\
    (0, 0, 3) & 0.2218 \\
    \hline\hline
    \end{tabular}
    \caption{\textbf{Pseudo-number resolved detection efficiencies}. The values reported are estimated according to the procedure described above.}
    \label{tab:efficiencies}
\end{table}

\subsection{Reconstruction of bunching probabilities with distinguishable photons}

To estimate the bunching probabilities in the fully distinguishable scenario for the random unitaries, we exploited that the 3-photon probabilities are a convex combination of single-particle ones. Given a unitary matrix $U$ describing the programmed interferometer, where the matrix element $U_{ij}$ describes the transmission amplitude between input mode $j$ and output mode $i$, and considering an input state with one-photon on each of the 3 input modes, the following relations hold:
\begin{eqnarray}
    p^{D}_{(3,0,0)} &=& \prod_{l=1}^3 \abs{U_{1l}}^2,\\ 
    p^{D}_{(0,3,0)} &=& \prod_{l=1}^3 \abs{U_{2l}}^2,\\ 
    p^{D}_{(0,0,3)} &=& \prod_{l=1}^3 \abs{U_{3l}}^2,
\end{eqnarray}
which depend only on the square moduli $\abs{U_{ij}}^2$ of the unitary matrix $U$. Such moduli can be measured by injected  in the IIP single photons into each input $j$, and measuring the corresponding counts $N_{ij}$ from each output mode $i$. Note that here the mode splitting matrix was not programmed on the device, thus resulting in a 1-to-1 correspondence between input and output modes. Then, the moduli square of the unitary matrix can be reconstructed as:
\begin{equation}
    T_{ij} = \abs{U_{ij}}^2 = \frac{N_{ij}}{\sum_i N_{ij}}.
\end{equation}

\subsection{Experimental data tables}

We conclude this section by describing the data analysis procedure supporting the data shown in Fig. \ref{fig:data_haar_3}. In particular, given a 3-mode unitary transformation $U$ programmed on the IIP, for the scenario related to the Gram matrix $S_A$, we proceeded by acquiring raw three-fold events $\tilde{N}_{(n_1,n_2,n_3)}$ at the output of the interferometric mesh, over a time span of $\sim 1$ hour, in the complete configuration composed by the unitary transformation plus the mode splitting setup. We then corrected such events as in Eq. \eqref{eqn:correction} to compute the experimental probability distribution $\tilde{p}_{(n_1,n_2,n_3)}$ for each output configuration $(n_1,n_2,n_3)$ as in Eq. \eqref{eqn:fullprob}. Then, we computed the bunching ratio $\tilde{r}_{FB}(S_A)$ as:
\begin{equation}
    \tilde{r}_{FB}(S_A) = \frac{\tilde{p}_{(3,0,0)} + \tilde{p}_{(0,3,0)} + \tilde{p}_{(0,0,3)}}{p^D_{(3,0,0)} + p^D_{(0,3,0)} + p^D_{(0,0,3)}} =
    \frac{\tilde{p}_{FB}(S_A)}{\tilde{p}_{FB}(S_D)},
\end{equation}
where the elements $p^D_{(n_1,n_2,n_3)}$ were reconstructed by measuring the moduli square of the unitary transformation $\abs{U_{ij}}^2$ as described above. In Table \ref{tab:fullbunchingAAA} we report the complete set of experimental data collected, related to the 23 chosen configurations of the three-mode unitary transformation $U$.

For the scenario described by the Gram matrix $S_{B'}$, that is a scenario in which one photon is made completely distinguishable from the others, the reconstruction of the probability distribution in such a scenario can be performed by injecting the interferometer with two photons in input ports $(1,3)$, and the measure the two photon probabilities $\{\tilde{p}_{(2,0,0)}, \tilde{p}_{(0,2,0)}, \tilde{p}_{(0,0,2)} \}$. To estimate the corresponding elements $\{\tilde{p}_{(3,0,0)}, \tilde{p}_{(0,3,0)}, \tilde{p}_{(0,0,3)} \}$ for this indistinguishability scenario, it is then enough to perform a classical convolution of the two-photon probabilities with the probabilities $T_{i2}$ obtained with a \textit{distinguishable} photon input in port $2$ according to:
\begin{eqnarray}
    \tilde{p}_{(3,0,0)} &=& \tilde{p}_{(2,0,0)} T_{12},\\ 
    \tilde{p}_{(0,3,0)} &=& \tilde{p}_{(0,2,0)} T_{22},\\ 
    \tilde{p}_{(0,0,3)} &=& \tilde{p}_{(0,0,2)} T_{32}.
\end{eqnarray}
The bunching ratio $\tilde{r}_{FB}(S_C)$ can be then estimated from these reconstructed probabilities according to the following expression:
\begin{equation}
\begin{aligned}
    \tilde{r}_{FB}(S_C) &= \frac{\tilde{p}_{(2,0,0)} T_{12} + \tilde{p}_{(0,2,0)} T_{22} + \tilde{p}_{(0,0,2)} T_{32}}{p^D_{(3,0,0)} + p^D_{(0,3,0)} + p^D_{(0,0,3)}} = \\
    &= \frac{\tilde{p}_{FB}(S_C)}{\tilde{p}_{FB}(S_D)}.
    \end{aligned}
\end{equation}
The complete set of data in this scenario $S_C$ with the 23 random unitaries is reported in Table \ref{tab:fullbunchingABA}.

\begin{table*}[h!]
    \begin{tabular}{c|c|c||ccc||ccc||ccc}
    \multirow{2}{*}{$\tilde{r}_{FB}(S_A)$} & \multirow{2}{*}{$\tilde{\sigma}(U_i)$} & \multirow{2}{*}{$\bar{\Delta}(U_i)$} &
    $\tilde{N}_{(3,0,0)}$ & \multirow{2}{*}{$\tilde{p}_{(3,0,0)}$} & \multirow{2}{*}{$\frac{\tilde{p}_{(3,0,0)}}{p^{D}_{(3,0,0)}}$} &
    $\tilde{N}_{(0,3,0)}$ & \multirow{2}{*}{$\tilde{p}_{(0,3,0)}$} & \multirow{2}{*}{$\frac{\tilde{p}_{(0,3,0)}}{p^{D}_{(0,3,0)}}$} &
    $\tilde{N}_{(0,0,3)}$ & \multirow{2}{*}{$\tilde{p}_{(0,0,3)}$} & \multirow{2}{*}{$\frac{\tilde{p}_{(0,0,3)}}{p^{D}_{(0,0,3)}}$}\\
    & & & $(\times 10^{3})$ & & & $(\times 10^{3})$ & & & $(\times 10^{3})$ & & \\
    \hline \hline
$4.84(9)$ & $0.736(3)$ & $-0.098(3)$ & $0.26(2)$ & $0.0138(8)$ & $4.9(3)$ & $1.58(4)$ & $0.083(2)$ & $4.9(1)$ & $0.47(2)$ & $0.025(1)$ & $4.7(2)$\\
$4.8(1)$ & $0.606(3)$ & $-0.170(2)$ & $0.92(3)$ & $0.048(2)$ & $4.7(1)$ & $0.34(2)$ & $0.0176(9)$ & $5.3(3)$ & $0.56(2)$ & $0.029(1)$ & $4.6(2)$\\
$4.77(6)$ & $0.857(4)$ & $-0.004(4)$ & $2.15(5)$ & $0.114(2)$ & $4.76(9)$ & $1.29(4)$ & $0.069(2)$ & $4.7(1)$ & $0.44(2)$ & $0.023(1)$ & $4.9(2)$\\
$4.7(1)$ & $0.711(4)$ & $-0.114(3)$ & $0.90(3)$ & $0.064(2)$ & $4.6(1)$ & $0.66(3)$ & $0.047(2)$ & $4.8(2)$ & $0.23(2)$ & $0.016(1)$ & $5.0(3)$\\
$4.87(5)$ & $1.071(5)$ & $0.289(9)$ & $2.15(5)$ & $0.145(3)$ & $5.1(1)$ & $2.30(5)$ & $0.156(3)$ & $4.78(9)$ & $0.89(3)$ & $0.060(2)$ & $4.6(1)$\\
$5.04(4)$ & $1.129(5)$ & $0.41(1)$ & $2.88(5)$ & $0.162(3)$ & $5.15(8)$ & $2.93(5)$ & $0.166(3)$ & $5.08(8)$ & $1.72(4)$ & $0.096(2)$ & $4.8(1)$\\
$4.99(7)$ & $0.889(4)$ & $0.030(5)$ & $0.94(3)$ & $0.053(2)$ & $4.8(2)$ & $1.97(4)$ & $0.113(2)$ & $4.9(1)$ & $0.89(3)$ & $0.051(2)$ & $5.3(2)$\\
$5.06(8)$ & $0.800(4)$ & $-0.048(4)$ & $1.79(4)$ & $0.103(2)$ & $5.0(1)$ & $0.85(3)$ & $0.049(2)$ & $5.0(2)$ & $0.48(2)$ & $0.027(1)$ & $5.2(2)$\\
$5.16(7)$ & $0.937(4)$ & $0.084(5)$ & $2.06(5)$ & $0.127(3)$ & $5.1(1)$ & $1.25(4)$ & $0.077(2)$ & $5.3(1)$ & $0.53(2)$ & $0.033(1)$ & $4.8(2)$\\
$4.83(3)$ & $1.163(5)$ & $0.51(1)$ & $3.57(6)$ & $0.171(3)$ & $4.99(7)$ & $3.61(6)$ & $0.175(3)$ & $4.88(7)$ & $3.00(5)$ & $0.144(2)$ & $4.60(8)$\\
$4.91(5)$ & $0.985(4)$ & $0.145(6)$ & $1.05(3)$ & $0.057(2)$ & $4.7(1)$ & $2.85(5)$ & $0.155(3)$ & $4.98(8)$ & $1.35(4)$ & $0.073(2)$ & $4.9(1)$\\
$5.2(1)$ & $0.766(3)$ & $-0.071(3)$ & $0.63(3)$ & $0.034(1)$ & $5.1(2)$ & $0.068(8)$ & $0.0037(4)$ & $13(2)$ & $1.38(4)$ & $0.074(2)$ & $5.1(1)$\\
$4.99(8)$ & $0.846(4)$ & $-0.010(4)$ & $1.53(4)$ & $0.082(2)$ & $5.2(1)$ & $1.55(4)$ & $0.084(2)$ & $4.7(1)$ & $0.10(1)$ & $0.0052(5)$ & $8.1(8)$\\
$4.91(4)$ & $1.101(5)$ & $0.349(9)$ & $2.87(5)$ & $0.152(3)$ & $5.16(9)$ & $2.20(5)$ & $0.118(2)$ & $4.87(9)$ & $2.76(5)$ & $0.147(2)$ & $4.70(8)$\\
$5.3(2)$ & $0.388(2)$ & $-0.248(1)$ & $0.12(1)$ & $0.0065(6)$ & $5.6(5)$ & $0.40(2)$ & $0.022(1)$ & $5.3(3)$ & $0.068(8)$ & $0.0037(4)$ & $5.3(6)$\\
$5.09(5)$ & $0.992(4)$ & $0.155(6)$ & $1.48(4)$ & $0.079(2)$ & $5.2(1)$ & $2.16(5)$ & $0.116(2)$ & $5.3(1)$ & $2.29(5)$ & $0.122(2)$ & $4.90(9)$\\
$5.0(1)$ & $0.693(3)$ & $-0.121(3)$ & $0.25(2)$ & $0.0136(9)$ & $5.1(3)$ & $1.02(3)$ & $0.056(2)$ & $5.2(2)$ & $0.92(3)$ & $0.050(2)$ & $4.7(1)$\\
$4.86(7)$ & $0.844(4)$ & $-0.013(4)$ & $2.30(5)$ & $0.120(2)$ & $4.79(9)$ & $0.21(1)$ & $0.0109(8)$ & $5.0(3)$ & $0.62(2)$ & $0.032(1)$ & $5.1(2)$\\
$5.09(7)$ & $0.830(4)$ & $-0.025(4)$ & $0.62(2)$ & $0.034(1)$ & $5.3(2)$ & $1.70(4)$ & $0.093(2)$ & $5.4(1)$ & $1.28(4)$ & $0.069(2)$ & $4.7(1)$\\
$4.87(4)$ & $1.031(4)$ & $0.216(7)$ & $1.04(3)$ & $0.055(2)$ & $5.3(2)$ & $3.00(5)$ & $0.158(3)$ & $4.95(8)$ & $2.21(5)$ & $0.116(2)$ & $4.57(9)$\\
$4.85(6)$ & $0.949(4)$ & $0.097(5)$ & $0.16(1)$ & $0.0086(7)$ & $8.3(6)$ & $1.27(4)$ & $0.067(2)$ & $5.3(1)$ & $3.10(6)$ & $0.163(3)$ & $4.58(7)$\\
$4.9(2)$ & $0.460(2)$ & $-0.232(1)$ & $0.036(6)$ & $0.0019(3)$ & $7(1) \times 10$ & $0.51(2)$ & $0.027(1)$ & $4.6(2)$ & $0.019(4)$ & $0.0010(2)$ & $4(1)$\\
$4.97(7)$ & $0.824(4)$ & $-0.030(4)$ & $1.57(4)$ & $0.079(2)$ & $5.0(1)$ & $0.19(1)$ & $0.0097(7)$ & $7.6(5)$ & $1.63(4)$ & $0.082(2)$ & $4.7(1)$\\
\hline
\hline
    \end{tabular}
    \caption{\textbf{Full table of the results obtained for the 23 unitaries for the scenario associated to the Gram matrix $S_A$.} Here, $\tilde{r}_{FB}(S_A) = \tilde{p}_{FB}(S_A)/\tilde{p}_{FB}(S_D)$ represents the experimentally measured overall bunching ratio $r_{FB}(S_A) = \text{per}(S_A)/\text{per}(S_D)$; while $\tilde{\sigma}(U_i)$ indicates the experimentally measured photon number variance at the output of the interferometer. We denote by $\tilde{N}_{(n_1,n_2,n_2)}$ the raw counts measured at the output of the pseudo number resolving detection setup.}
    \label{tab:fullbunchingAAA}
\end{table*}

\begin{table*}[h!]
   \centering
    \begin{tabular}{c||ccc||ccc||ccc}
    \multirow{2}{*}{$\tilde{r}_{FB}(S_C)$} & 
    $\tilde{N}_{(2,0,0)}$ & \multirow{2}{*}{$\tilde{p}_{(2,0,0)} T_{12}$} & \multirow{2}{*}{$\frac{\tilde{p}_{(3,0,0)}}{p^{D}_{(3,0,0)}}$} &
    $\tilde{N}_{(0,2,0)}$ & \multirow{2}{*}{$\tilde{p}_{(0,2,0)} T_{22}$} & \multirow{2}{*}{$\frac{\tilde{p}_{(0,3,0)}}{p^{D}_{(0,3,0)}}$} &
    $\tilde{N}_{(0,0,2)}$ & \multirow{2}{*}{$\tilde{p}_{(0,0,2)} T_{32}$} & \multirow{2}{*}{$\frac{\tilde{p}_{(0,0,3)}}{p^{D}_{(0,0,3)}}$}\\
    & $(\times 10^{4})$ & & & $(\times 10^{4})$ & & & $(\times 10^{4})$ & & \\
    \hline \hline
$1.83(2)$ & $1.25(1)$ & $0.00510(4)$ & $1.80(1)$ & $0.432(7)$ & $0.0309(5)$ & $1.80(3)$& $0.272(5)$ & $0.0102(2)$ & $1.94(4)$\\
$1.77(1)$ & $1.87(1)$ & $0.0184(1)$ & $1.80(1)$ & $0.056(2)$ & $0.0055(2)$ & $1.65(7)$& $2.36(2)$ & $0.01115(6)$ & $1.79(1)$\\
$1.77(1)$ & $1.30(1)$ & $0.0400(3)$ & $1.67(1)$ & $0.354(6)$ & $0.0281(5)$ & $1.93(3)$& $2.77(2)$ & $0.00847(4)$ & $1.802(9)$\\
$1.71(2)$ & $0.717(8)$ & $0.0211(2)$ & $1.52(2)$ & $0.238(5)$ & $0.0189(4)$ & $1.89(4)$& $1.51(1)$ & $0.00616(4)$ & $1.90(1)$\\
$1.814(6)$ & $2.68(2)$ & $0.0490(2)$ & $1.728(9)$ & $2.31(2)$ & $0.0583(3)$ & $1.78(1)$& $0.405(6)$ & $0.0270(4)$ & $2.09(3)$\\
$1.785(7)$ & $1.10(1)$ & $0.0604(5)$ & $1.92(2)$ & $1.24(1)$ & $0.0548(5)$ & $1.68(1)$& $3.09(2)$ & $0.0351(2)$ & $1.743(8)$\\
$1.776(9)$ & $0.95(1)$ & $0.0190(2)$ & $1.71(2)$ & $2.08(1)$ & $0.0397(2)$ & $1.74(1)$& $0.264(5)$ & $0.0187(4)$ & $1.95(4)$\\
$1.75(1)$ & $0.467(7)$ & $0.0336(5)$ & $1.65(2)$ & $1.31(1)$ & $0.0179(1)$ & $1.83(1)$& $0.403(6)$ & $0.0106(2)$ & $2.02(3)$\\
$1.86(1)$ & $0.701(8)$ & $0.0461(5)$ & $1.87(2)$ & $0.669(8)$ & $0.0276(3)$ & $1.91(2)$& $2.49(2)$ & $0.01175(6)$ & $1.736(9)$\\
$1.766(5)$ & $1.39(1)$ & $0.0596(5)$ & $1.74(1)$ & $1.75(1)$ & $0.0646(4)$ & $1.81(1)$& $1.82(1)$ & $0.0547(4)$ & $1.75(1)$\\
$1.782(9)$ & $2.16(1)$ & $0.0211(1)$ & $1.75(1)$ & $1.05(1)$ & $0.0554(5)$ & $1.78(2)$& $0.572(8)$ & $0.0270(3)$ & $1.82(2)$\\
$1.81(2)$ & $0.166(4)$ & $0.0109(3)$ & $1.66(4)$ & $1.33(1)$ & $0.000480(4)$ & $1.75(1)$& $0.606(8)$ & $0.0269(3)$ & $1.87(2)$\\
$1.83(2)$ & $0.427(7)$ & $0.0298(4)$ & $1.87(3)$ & $0.747(9)$ & $0.0322(4)$ & $1.81(2)$& $2.10(1)$ & $0.001088(6)$ & $1.69(1)$\\
$1.743(6)$ & $0.812(9)$ & $0.0480(5)$ & $1.62(2)$ & $2.61(2)$ & $0.0452(2)$ & $1.873(9)$& $1.51(1)$ & $0.0548(4)$ & $1.75(1)$\\
$1.84(2)$ & $1.78(1)$ & $0.00206(1)$ & $1.78(1)$ & $1.52(1)$ & $0.00734(5)$ & $1.78(1)$& $0.015(1)$ & $0.0016(1)$ & $2.3(2)$\\
$1.851(8)$ & $1.12(1)$ & $0.0299(3)$ & $1.95(2)$ & $1.85(1)$ & $0.0403(3)$ & $1.83(1)$& $0.726(9)$ & $0.0449(5)$ & $1.81(2)$\\
$1.78(2)$ & $1.49(1)$ & $0.00499(4)$ & $1.88(1)$ & $0.238(5)$ & $0.0192(4)$ & $1.78(4)$& $0.656(8)$ & $0.0184(2)$ & $1.75(2)$\\
$1.760(9)$ & $2.76(2)$ & $0.0421(2)$ & $1.675(8)$ & $0.066(3)$ & $0.0062(2)$ & $2.8(1)$& $3.47(2)$ & $0.01102(4)$ & $1.730(7)$\\
$1.91(1)$ & $1.13(1)$ & $0.0117(1)$ & $1.83(2)$ & $1.03(1)$ & $0.0330(3)$ & $1.91(2)$& $0.425(7)$ & $0.0289(4)$ & $1.94(3)$\\
$1.788(6)$ & $0.242(5)$ & $0.0176(4)$ & $1.72(3)$ & $2.63(2)$ & $0.0596(3)$ & $1.865(9)$& $2.71(2)$ & $0.0434(2)$ & $1.717(8)$\\
$1.838(9)$ & $0.050(2)$ & $0.0028(1)$ & $2.7(1)$ & $1.51(1)$ & $0.0253(2)$ & $2.01(1)$& $1.56(1)$ & $0.0622(4)$ & $1.75(1)$\\
$1.99(3)$ & $0.0041(6)$ & $0.00037(6)$ & $14(2)$ & $0.490(7)$ & $0.0113(2)$ & $1.94(3)$& $0.460(7)$ & $0.000423(6)$ & $1.81(3)$\\
$1.819(6)$ & $2.66(2)$ & $0.0290(1)$ & $1.841(9)$ & $0.029(2)$ & $0.0027(2)$ & $2.1(1)$& $3.03(2)$ & $0.0309(1)$ & $1.778(8)$\\
\hline
\hline
    \end{tabular}
    \caption{\textbf{Full table of the results obtained for the 23 unitaries, for the scenario associated to the Gram matrix $S_C$.} Here, $\tilde{r}_{FB}(S_C) = \tilde{p}_{FB}(S_C)/\tilde{p}_{FB}(S_D)$ represents the experimentally measured overall bunching ratio $r_{FB}(S_C) = \text{per}(S_C)/\text{per}(S_D)$. We denote by $\tilde{N}_{(n_1,n_2,n_3)}$ the raw counts measured at the output of the pseudo number resolving detection setup, for the two-photon input as described above. The three-photon bunching probabilities can be then retrieved as $\{\tilde{p}_{(n_1,n_2,n_3)} T_{12}, \tilde{p}_{(n_1,n_2,n_3)} T_{22}, \tilde{p}_{(n_1,n_2,n_3)} T_{32}\}$, where $T_{ij}$ represents the measured square moduli $\abs{U_{ij}}^2$ of the programmed unitary matrix.}
    \label{tab:fullbunchingABA}
\end{table*}

%

\end{document}